\documentclass[preprint,12pt,authoryear]{elsarticle}

\usepackage{amssymb}
\usepackage{algorithm}
\usepackage{algorithmic}
\usepackage{float}
\usepackage{booktabs}
\usepackage{diagbox}
\usepackage{multirow}
\usepackage{amsmath, amsfonts,latexsym,amsthm, amsxtra,graphicx,color}
\usepackage{url}

\newtheorem{theorem}{Theorem}
\newtheorem{lemma}{Lemma}
\newproof{pf}{Proof}
\newproof{pot1}{Proof of Theorem~\ref{thm1}}
\newproof{pot2}{Proof of Theorem~\ref{thm2}}
\newproof{poc1}{Proof of Corollary~\ref{crl1}}
\newproof{poc2}{Proof of Corollary~\ref{crl2}}
\newdefinition{rmk}{Remark}

\newtheorem{crl}{Corollary}
\newdefinition{case}{Case}
\newdefinition{assumption}{Assumption}

\begin{document}

\begin{frontmatter}



\title{A novel approach of empirical likelihood with massive data}


\author[inst1]{Yang Liu}

\affiliation[inst1]{organization={School of Mathematics and Statistics},
            addressline={Shaanxi Normal University}, 
            city={Xi'an},
            postcode={710119}, 
            country={China}}

\author[inst1]{Xia Chen}
\author[inst1]{Wei-min Yang}


\begin{abstract}
In this paper, we propose a novel approach for tackling the obstacles of empirical likelihood in the face of massive data, which is called split sample mean empirical likelihood (SSMEL), our approach provides a unique perspective for solving big data problems. We show that the SSMEL estimator has the same estimation efficiency as the empirical likelihood estimator with the full dataset, and maintains the important statistical property of Wilks' theorem, allowing our proposed approach to be used for statistical inference without estimating the covariance matrix. This effectively tackles the hurdle of the Divide and Conquer (DC) algorithm for statistical inference. We further illustrate the proposed approach via simulation studies and real data analysis.
\end{abstract}



\end{frontmatter}


\section{Introduction}
As science and technology continue to advance, datasets are growing in size at an accelerated rate, making large datasets increasingly common. For example, Barclaycard (UK) carries out 350 million transactions a year, Wal-Mart makes over 7 billion transactions a year, and AT\& T carries over 70 billion long-distance calls annually \citep{Hand2000}. The abundance of massive data presents new challenges for classical statistical methods. While these methods may offer excellent theoretical properties for analyzing such data, they can be difficult to implement in practice due to constraints on computation time and memory. Moreover, storing data in a distributed manner can make it impractical to conduct statistical analysis on the entire dataset due to communication costs and privacy issues. Consequently, there is an increasing demand for a novel statistical approach to tackle the difficulties posed by enormous data. Recently, Divide and Conquer (DC) have gained widespread popularity for addressing the issues related to massive data. DC algorithms are effective in statistical analysis problems with massive data. Many studies, including \citet{lin2011aggregated,chen2014split,lee2017,battey2018distributed,shi2018massive,fan2019distributed,
chen2019quantile,jordan2019,fan2021communication,chen2021distributed,chen2021aos} have successfully utilized the one-shot approach and the iterative approach of DC algorithms in various statistical models. We refer to \citet{gao2022review} for a recent review of distributed statistical literature. While DC algorithms have proven to be useful in parameter estimation, statistical inference remains a complex task within this framework. The general statistical inference methods rely on the asymptotic distribution of estimators to determine the test statistic. However, estimating the covariance matrix under distributed or massive data can be challenging.

Empirical likelihood is a significant nonparametric and semiparametric statistical method, it holds Wilks' theorem of parametric likelihood \citep{owen88,qin1994}. Therefore, it produces confidence regions with data-driven shapes and constructs test statistics without estimating the covariance matrix. \citet{DiCiccio91} demonstrated that empirical likelihood resembles parametric likelihood with Bartlett correction. Due to these advantageous properties, and empirical likelihood can easily incorporate side information, so it has gained significant attention and has been extensively investigated and utilized, e.g. regression models \citep{owen91,chen2009a}, estimating equations \citep{qin1994}, partially linear models \citep{shi2000}, bayesian settings \citep{lazar2003}, quantile regression models \citep{whang2006,otsu2008}, U-statistics \citep{jing2009}, time series models \citep{kitamura1997,chen2003}, high-dimensional statistical inference \citep{hjort2009,chen2009b,leng2012,chang2018,chang2021}.

Empirical likelihood can be computationally intensive, particularly when dealing with large datasets, which can limit its applicability. Because empirical likelihood is well linked to traditional statistical models and has a unique advantage in statistical inference, it is essential to overcome these challenges when working with massive data. Recently, \citet{ssel2020} and \citet{del2022} proposed split sample empirical likelihood (SSEL) and distributed empirical likelihood (DEL), respectively, to solve this problem. \citet{ssel2020} constructed the empirical likelihood function concerning each subset and defined the SSEL estimator as the maximizer of the product of these empirical likelihood functions. More extensive works based on this idea can be found in \citet{zhou2022distributed}. \citet{del2022} obtained the estimators for each subset and then averaged these estimators across all subsets to generate the DEL estimator. Both methods utilize parallel computing to tackle the challenges of massive data on empirical likelihood. Modern parallel computing structures have the potential to significantly reduce computation time. However, for large split size $K$ (exceeding $o(n^{1/2})$), the accuracy of estimation, particularly for non-linear models, cannot be ensured. Consequently, there are stringent limitations on the value of $K$ required to obtain reliable estimators, and each parallel pool remains computationally expensive. On the other hand, the DEL is failing to meet Wilks' theorem, which eliminates the benefits of empirical likelihood and necessitates the exploration of alternative statistical inference methods.

To address these issues, we propose a novel approach, which is called the split sample mean empirical likelihood (SSMEL). Under mild regularity conditions, we show that the SSMEL estimator retains the same asymptotic efficiency as that of the full dataset, and it holds the important property of Wilks' theorem.  Our investigation contributes to several areas. First, empirical likelihood offers a wide range of applications since it has been widely extended to conventional statistical models including linear models, quantile regression, U-statistics, and so on. Our approach successfully addresses the challenge of empirical likelihood caused by massive data and broadens the scope of empirical likelihood. Second, the SSMEL solves the dilemma of empirical likelihood with big data without using parallel structures, so it can be implemented efficiently with general computing devices, making it more widely practical and applicable.
Finally, statistical inference using empirical likelihood offers unique benefits since it does not need to estimate the covariance matrix. Wilks' theorem holds for the SSMEL, making statistical inference using the SSMEL easy and efficient when dealing with massive data. In addition, we expand the algorithm in \citet{tang2014algo} to support the SSMEL for distributed data, which is a variant of the iterative approach. 

The rest of this paper is organized as follows. In Section 2, we briefly review empirical likelihood and present the methodology of the SSMEL. Section 3 investigates the theoretical properties of the SSMEL. Section 4 designs a new algorithm applicable to the SSMEL. Sections 5 and 6 examine the performance of the proposed approach on simulated and real data analysis. Section 7 concludes the paper and discusses future work.

\section{Methodology}
Suppose that $\mathcal{X} = \{x_{1},\cdots,x_{n}\}$ are $d$-variate independent and identically distributed samples with common distribution function $F$. Let $\theta\in\mathbb{R}^{p}$ be a vector of the unknown parameter of interest, and $\theta_0$ is the true value. For the sake of completeness, we first briefly review the empirical likelihood.

\subsection{Empirical likelihood}

Assume that the truth value $\theta_0$ satisfies constraints in the form of the $r\ge p$ unbias estimating equation, i.e.
\begin{equation*}
 \mathbb{E}g(X,\theta_0)=0,   
\end{equation*}
where $g(X,\theta_0)=(g_1(X,\theta_0),\cdots,g_r(X,\theta_0))$. Then, the empirical likelihood ratio function evaluated at $\theta$ can be defined as
\begin{equation}
\label{el}
R(\theta)=\sup\left\{\prod\limits_{i=1}\limits^{n}np_{i}:p_{i}\ge0,\sum\limits_{i=1}\limits^{n}p_{i}=1,\sum\limits_{i=1}\limits^{n}p_{i}g(x_{i},\theta)=0\right\}.
\end{equation}
By the Lagrange multiplier method, we have
\begin{equation*}
p_{i}=\frac{1}{n}\cdot\frac{1}{1+\lambda^{T}g(x_{i},\theta)},
\end{equation*}
where $\lambda(\theta)$ is the solution to following equations: 
\begin{equation*}
\frac{1}{n}\sum\limits_{i=1}\limits^{n}\frac{g(x_{i},\theta)}{1+\lambda^{T}g(x_{i},\theta)}=0.
\end{equation*}
Thus, the empirical log-likelihood ratio function for $\theta$ is given by
\begin{equation}
\label{logel}
\ell(\theta)=\sum\limits_{i=1}\limits^{n}\log\left[1+\lambda^{T}g(x_{i},\theta)\right]
\end{equation}
The maximum empirical likelihood estimator $\hat{\theta}_{EL}$ is calculated by
\begin{equation}
\label{el_est}
\hat{\theta}_{EL}=\arg\min\limits_{\theta\in\Theta}\max\limits_{\lambda\in\hat{\Lambda}_{n}(\theta)}\sum\limits_{i=1}\limits^{n}\log\left[1+\lambda^{T}g(x_{i},\theta)\right],
\end{equation}
where $\hat{\Lambda}_{n}(\theta)=\{\lambda\in\mathbb{R}^{r}:\lambda^{T}g(x_{i},\theta)\in\mathcal{V},i=1,\cdots,n\}$for $\theta\in\Theta$ and $\mathcal{V}$ is an open interval containing zero, and $\Theta$ is the convex hull of $\{g(x_{i},\theta),i=1\cdots,n)\}$. Under mild regularity conditions, \citet{qin1994} showed  that as $n\rightarrow\infty$,
\begin{equation*}
\sqrt{n}\left(\hat{\theta}_{EL}-\theta_0\right)\stackrel{d}{\longrightarrow}N(0,\Sigma),
\end{equation*}
where
\begin{equation*}
\Sigma=\left[\mathbb{E}\big(\frac{\partial g(X,\theta_0)}{\partial\theta^{T}}\big)^{T}(\mathbb{E}g(X,\theta_0)g^{T}(X,\theta_0))^{-1}\mathbb{E}\big(\frac{\partial g(X,\theta_0)}{\partial\theta^{T}}\big)\right]^{-1}.
\end{equation*}
Moreover, if $Var\left(g(X,\theta_0)\right)$ is finite and the rank $p>0$, then Wilks' theorem is hold, i.e.
\begin{equation*}
2\ell(\theta_0)-2\ell(\hat{\theta}_{EL})\stackrel{d}{\longrightarrow}\chi_{p}^{2},\ as\  n\rightarrow\infty.
\end{equation*}

\subsection{Split sample mean empirical likelihood}

Empirical likelihood cannot generally be written in a closed form, so a numerical optimization algorithm is required for the solution, resulting in computational obstacles for massive and distributed data. To address these challenges, we introduce our proposed SSMEL. Assume the size of full dataset $n$ is very large, and randomly partition the full dataset $\mathcal{X}=\{x_{1},\cdots,x_{n}\}$ into $K$ subsets of size $m=n/K$. We denote $S_{k}=\{x_{i}^{(k)},i=1,\cdots,m\}$ as $k$th subset, which $x_{i}^{(k)}$ means $i$th sample in $k$th subset. Obviously, $\bigcup_{k=1}^{K}S_{k}=\mathcal{X}$ and $S_{k}\bigcap S_{t}=\emptyset$, for any $k\neq t$.

For each subset $S_{k},k=1\cdots,K$, we consider the following steps:
\begin{itemize}
\item By inputting the samples $x_{i}^{(k)}$ from subset $S_{k}$ into the estimating function, we can obtain the sequence of estimating functions $g(x_{i}^{(k)},\theta),i=1,\cdots,m$.

\item Taking the mean of estimating function sequence $\left\{g(x_{1}^{(k)},\theta),\cdots,g(x_{m}^{(k)},\theta)\right\}$, i.e.
\begin{equation*}
\bar{g}^{(k)}(\theta)=\frac{1}{m}\sum\limits_{i=1}\limits^{m}g(x^{(k)}_{i},\theta).    
\end{equation*}
\end{itemize} 

With the aforementioned steps, we can derive the mean estimating functions $\bar{g}^{(k)}(\theta),k=1,\cdots, K$ from $K$ subsets. It can be easily seen that the mean estimating function is still the estimating equation, i.e.
\begin{equation*}
\mathbb{E}\bar{g}(\theta_0)=\mathbb{E}\left[\frac{1}{m}\sum\limits_{i=1}\limits^{m}g(X_{i},\theta_0)\right]=\frac{1}{m}\sum\limits_{i=1}\limits^{m}\mathbb{E}g(X_{i},\theta_0)=0.    
\end{equation*}

Thus, we can construct the split sample mean empirical likelihood (SSMEL) ratio function using the mean estimating equations,
\begin{equation}
\label{ssmel}
R_{S}(\theta)=\sup\left\{\prod\limits_{k=1}\limits^{K}Kp_{k}:p_{k}\ge0,\sum\limits_{k=1}\limits^{K}p_{k}=1,\sum\limits_{k=1}\limits^{K}p_{k}\bar{g}^{(k)}(\theta)=0\right\},
\end{equation}
and the split sample mean empirical log-likelihood ratio function is
\begin{equation}
\label{logssmel}
\ell_{S}(\theta)=\sum\limits_{k=1}\limits^{K}\log\left[1+\lambda^{T}\bar{g}^{(k)}(\theta)\right].
\end{equation}

The core idea of the SSEL and DEL is to split up large-scale datasets into several smaller datasets utilizing parallel structures for simultaneous processing, which is a solution to the computational issues that arise from extremely large sample sizes in empirical likelihood. To accomplish this, high-quality computing equipment is needed. The fundamental goal of our approach is to compress the information provided in the estimating equations to directly transform intolerably massive samples into tolerably tiny samples. Thus, the full dataset empirical likelihood is a special case of the SSMEL, when $K=n$.

\begin{figure}
    \begin{minipage}[t]{0.49\linewidth}
        \centering
        \includegraphics[width=0.75\textwidth]{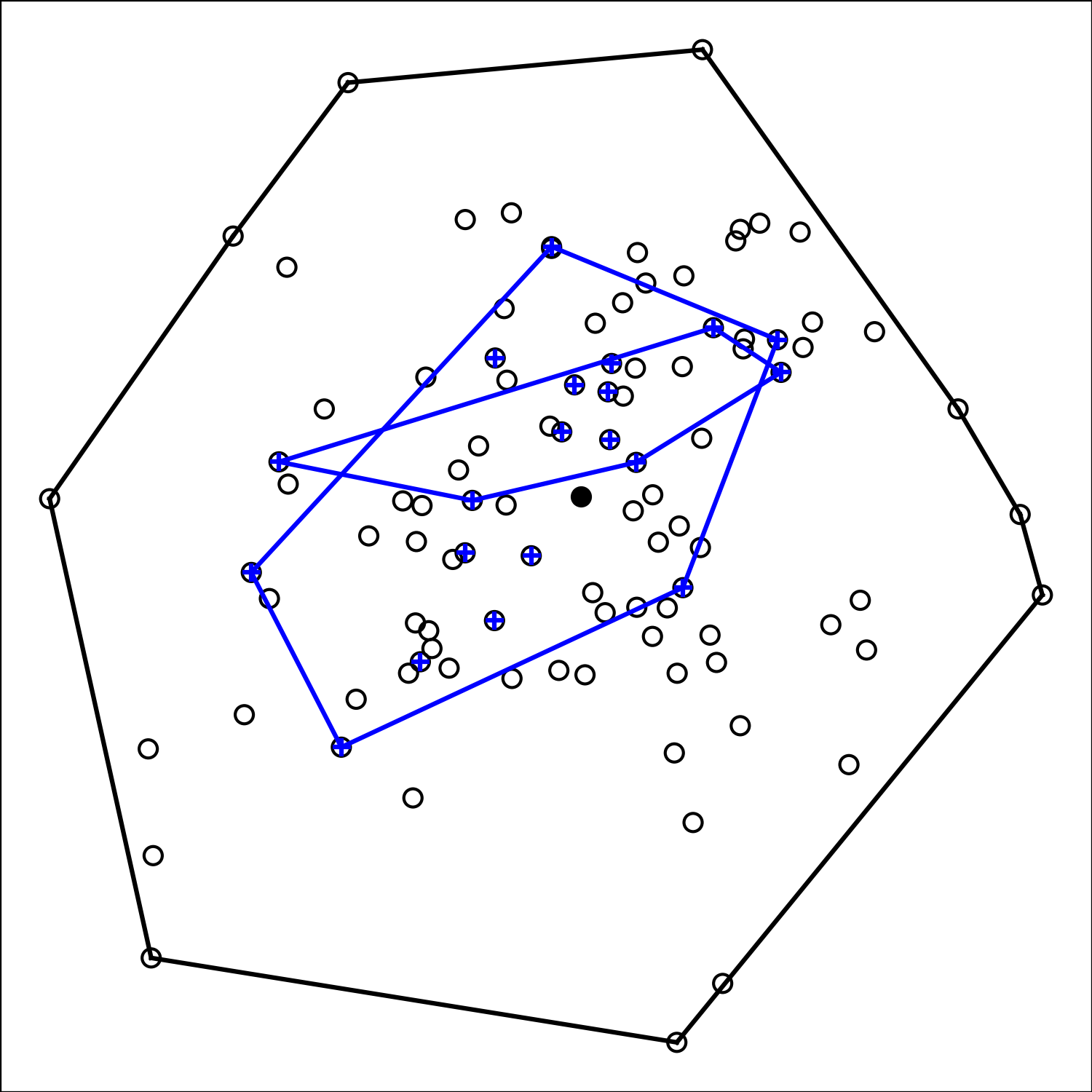}
        \centerline{(a)}
    \end{minipage}
 \begin{minipage}[t]{0.49\linewidth}
        \centering
        \includegraphics[width=0.75\textwidth]{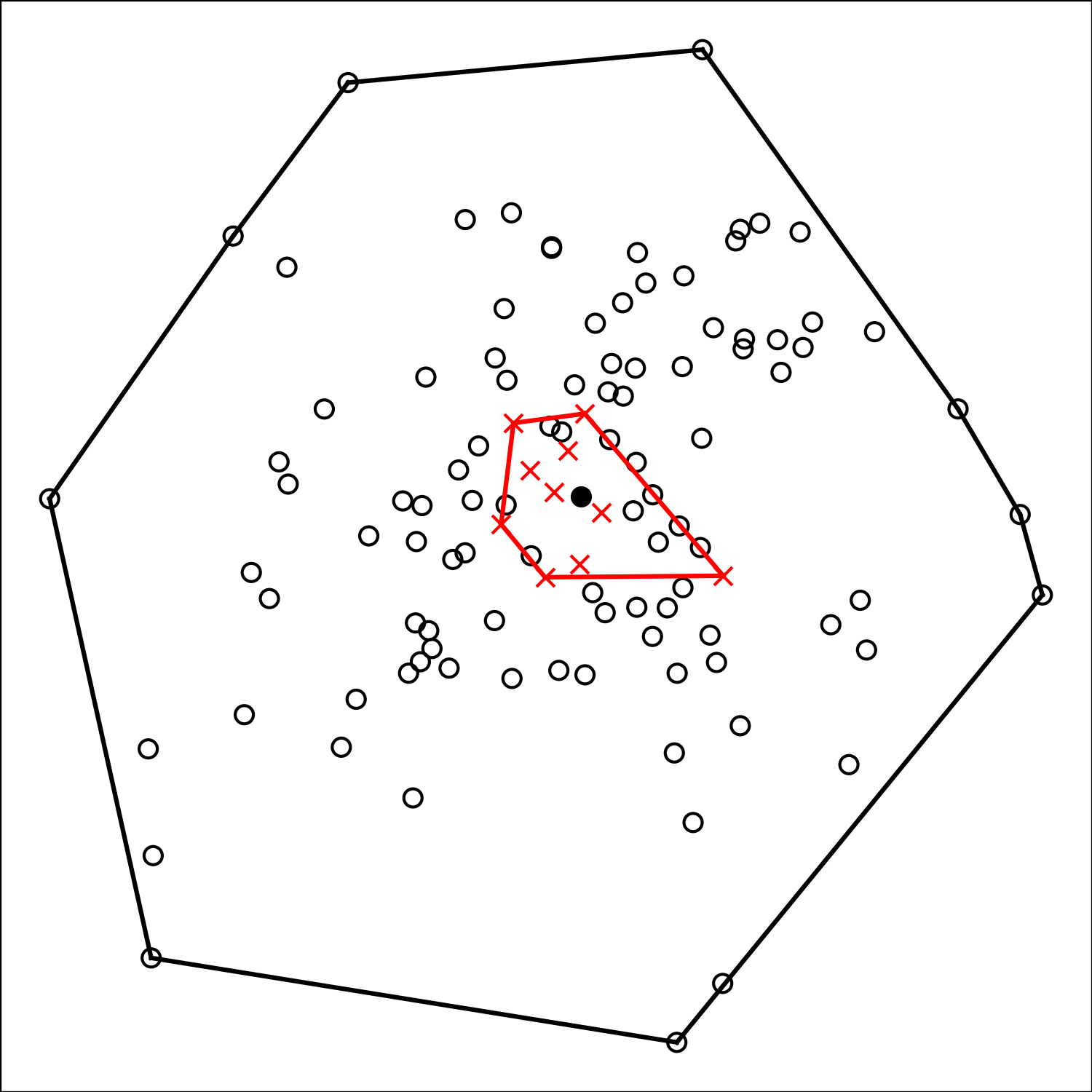}
        \centerline{(b)}
    \end{minipage}
    \caption{The black line represents the convex hull of the full dataset. The blue line in Fig.~\protect\ref{param_space}(a) shows the convex hull of partly subsets. The red line in Fig.~\protect\ref{param_space}(b) shows the convex hull of the SSMEL. The black solid dot represents the zero vectors.}\label{param_space}
\end{figure}

Similar to Equation~(\ref{el_est}), the maximum SSMEL estimator is
\begin{equation}
\label{ssmel_est}
\hat{\theta}_{S}=\arg\min\limits_{\theta\in\bar{\Theta}}\max\limits_{\lambda\in\hat{\Lambda}_{K}(\theta)}\sum\limits_{k=1}\limits^{K}\log\left[1+\lambda^{T}\bar{g}^{(k)}(\theta)\right],
\end{equation}
where $\hat{\Lambda}_{K}(\theta)=\big\{\lambda:\lambda^{T}\bar{g}^{(k)}(\theta)\in\mathcal{V},k=1,\cdots,K\big\}$ for $\theta\in\bar{\Theta}$ and $\mathcal{V}$ is an open interval containing zero, and $\bar{\Theta}$ is the convex hull of $\{\bar{g}^{(k)}(\theta),k=1,\cdots,K\}$. To solve Equation~(\ref{ssmel_est}), a prerequisite is that $\bar{\Theta}$ has the zero vector as an interior point. Lemma 11.1 in \citet{owen2001} states that if $Var\left(g(X,\theta)\right)$ is finite and the rank $p>0$, then the zero vector must be contained in $\Theta$. Obviously, since $Var\left(\bar{g}(\theta)\right)=m^{-1}Var\left(g(X,\theta)\right)$, if $Var\left(g(X,\theta)\right)$ satisfies this condition, then $Var\left(\bar{g}(\theta)\right)$ also satisfies it. Fig.~\protect\ref{param_space} shows the parameter space for partly subsets and the parameter space for the SSMEL under the same segmentation. It can be seen from Fig.~\protect\ref{param_space}(a) that when the subset size $m$ is small, the convex hull consisting of $g(x_i^{(k)},\theta),x_i^{(k)}\in S_{k}$ does not contain zero vectors with a higher probability, thus leading to poor estimation of the SSEL and DEL when the value of $K$ is taken to be large. In contrast, Fig.~\protect\ref{param_space}(b) shows that the SSMEL has a much smaller parameter space and always contains zero vectors.

\section{Asymptotic properties}
In this section, we establish the asymptotic properties of the SSMEL. For the empirical likelihood, the critical aspect is to control the tail probabilities behavior of the estimating function, i.e., to ensure $\left\Vert n^{-1}\sum_{i=1}^{n}g(x_{i},\theta)\right\Vert=O_{p}(n^{-1/2})$. It is worth noting that 
\begin{equation*}
K^{-1}\sum_{k=1}^{K}\bar{g}^{(k)}(\theta)=K^{-1}\sum_{k=1}^{K}m^{-1}\sum_{i=1}^{m}g(x_{i}^{(k)},\theta)=n^{-1}\sum_{i=1}^{n}g(x_{i},\theta).
\end{equation*} 
Thus,
\begin{equation*} 
\left\Vert K^{-1}\sum_{k=1}^{K}\bar{g}^{(k)}(\theta)\right\Vert=O_{p}\left(n^{-1/2}\right),
\end{equation*}
the SSMEL and empirical likelihood have the same assumptions. The following assumptions are made.
\begin{assumption}\label{A1}
$\theta_0\in int(\bar{\Theta})$ is unique solution to $\mathbb{E}g(X,\theta)=0$, where $\bar{\Theta}$ is a compact set and $int(\bar{\Theta})$ denotes the interior of $\bar{\Theta}$.
\end{assumption}
\begin{assumption}\label{A2}
$g(x_{i},\theta)$ is continuous with respect to $\theta$ at each $\theta\in\bar{\Theta}$ with probability 1 and is continuously differentiable with respect to $\theta$ in a neighbourhood of $\mathcal{N}$ of $\theta_0$.
\end{assumption}
\begin{assumption}\label{A3}
$\mathbb{E}\left[\sup\limits_{\theta\in\Theta}\left\Vert g(X,\theta)\right\Vert^\alpha\right]<\infty$ for some $\alpha>2$, where $\Vert\cdot\Vert$ is the Euclidean norm for vector and the Frobenius norm for matrix.
\end{assumption}
\begin{assumption}\label{A4}
$\Omega:=\mathbb{E}\left[g(X,\theta_0)g(X,\theta_0)^{T}\right]$ is nonsingular.
\end{assumption}
\begin{assumption}\label{A5}
$\mathbb{E}\left[\sup\limits_{\theta\in\mathcal{N}}\left\Vert \partial g(X,\theta)/\partial\theta^T\right\Vert\right]<\infty$, denotes $\mathbb{E}\left(\partial g(X,\theta)/\partial\theta^T\right)=G$, $rank(G)=p$.
\end{assumption}

\begin{rmk}\label{rmk1}
Assumptions~\protect\ref{A1}-\protect\ref{A5} guarantee the existence and asymptotic normality of $\hat{\theta}_{S}$, further ensure Wilks' theorem holds. As discussed in the previous section, the assumption~\protect\ref{A1} of $\bar{\Theta}$ can be relaxed to $\Theta$. These assumptions are similar to those in \citet{newey2004higher}, which are the fundamental assumptions of empirical likelihood, and no additional assumptions are introduced in this paper.
\end{rmk}

\begin{theorem}
\label{thm1}
Under the Assumptions~{\rm \protect\ref{A1}-\protect\ref{A5}}, we have
\begin{equation*}
\sqrt{n}\left(\hat{\theta}_{S}-\theta_0\right)\stackrel{d}{\longrightarrow}N(0,\Sigma),\ as\ n\rightarrow\infty,
\end{equation*}
where $\Sigma=\left(G^{T}\Omega G\right)^{-1}$.
\end{theorem}

\begin{crl}\label{crl1}
Under the assumptions of Theorem~{\rm\protect\ref{thm1}}, we have
\begin{equation*}
\mathbb{E}\left[\left\Vert\hat{\theta}_{S}-\theta_0\right\Vert^2\right]\le\frac{tr(\Sigma)}{n}+o\left(n^{-1}\right),
\end{equation*}
where $tr(\cdot)$ represents the trace of the matrix.
\end{crl}

Theorem~\protect\ref{thm1} shows that the asymptotic distribution of $\hat{\theta}_{S}$ is the same as for $\hat{\theta}_{EL}$, and if the estimating function $g$ is the score function of the true parameter likelihood function, then the asymptotic distribution of $\hat{\theta}_{S}$ is same as maximum likelihood estimator. Corollary~\protect\ref{crl1} shows that the mean squared error (MSE) upper bound for the SSMEL estimator, which is the same as the full dataset empirical likelihood, therefore they have the same estimation efficiency. Next, we give the asymptotic behavior of the SSMEL test statistic. Theorem~\protect\ref{thm2} summarizes the general conclusions, while Corollary~\protect\ref{crl2} provides the asymptotic distribution in the presence of nuisance parameters.

\begin{theorem}
\label{thm2}
The SSMEL ratio test statistic for $H_0:\theta=\theta_0$ is
\begin{equation*}
\mathcal{W}(\theta_0)=2\left[\ell_{S}(\theta_0)-\ell_{S}(\hat{\theta}_{S})\right].
\end{equation*}
Under the assumptions of  Theorem~{\rm\protect\ref{thm1}}, $\mathcal{W}(\theta_0)\stackrel{d}{\longrightarrow}\chi^{2}_{p} \ as \ n\rightarrow\infty$, when $H_0$ is true.
\end{theorem}

\begin{crl}\label{crl2}
Let $\theta^{T}=(\phi,\gamma)^{T}$, and $\phi$ is $q\times1$ vector, $\gamma$ is $(p-q)\times1$ nuisance parameters. The profile SSMEL ratio test statistic for $H_0:\phi=\phi_0$ is
\begin{equation*}
\mathcal{W}(\phi_0)=2\left[\ell_{S}\left(\phi_0,\hat{\gamma}(\phi_0)\right)-\ell_{S}\left(\hat{\phi}_S,\hat{\gamma}_S\right)\right].
\end{equation*}
where $\hat{\gamma}(\phi_0)$ minimizes $\ell_{S}(\phi_0,\gamma)$ with respect to $\gamma$. Under the assumptions of Theorem~{\rm\protect\ref{thm1}}, $\mathcal{W}(\phi_0)\stackrel{d}{\longrightarrow}\chi^{2}_{q}$ as $n\rightarrow\infty$, when $H_0$ is true.
\end{crl}

\begin{rmk}
Related to the choice of $K$, there are some considerations. As we formally use $K$ samples, the computation time grows as $K$ increases, therefore it is necessary to ensure that $K$ is not excessively large. On the other hand, the empirical likelihood can only be applied when $K$ is larger than $p$ (the parameter dimension). We advise a value of $K$ of at least 100 to ensure numerical convergence based on our experience. In the subsequent simulations, it was discovered that the SSMEL has a higher computational efficiency when $K$ is over 100 and has been able to be compatible with the full-sample empirical likelihood results.
\end{rmk}

\section{Algorithm for distributed data}
The algorithms for solving empirical likelihood can be applied to the SSMEL, implementing the SSMEL estimation feasible on a single computing device. To extend the SSMEL to the distributed data, we generalize the two-layer coordinate descent algorithm in \citet{tang2014algo}. The algorithm is briefly reviewed in the context of the SSMEL.

First, we define
\begin{equation}\label{inner_layer}
f(\lambda;\theta) = \frac{1}{K}\sum\limits_{k=1}^{K}\log_{*}\left\{1+\lambda^{T}\bar{g}^{(k)}(\theta)\right\}
\end{equation}
\begin{equation}\label{outer_layer}
 f(\theta) = \max\limits_{\lambda\in\hat{\Lambda}_{K}(\theta)}f(\lambda;\theta)
\end{equation}
where $\log_{*}(x)$ is a pseudo-logarithm function that is twice differentiable and has bounded support adopted from \citet{owen2001}:
\begin{equation*}
 \log_{*}(x) = \left\{
\begin{array}{ll}
\log(x)      &\quad       if\ x\ge\varepsilon\\
\log(\varepsilon)-1.5+2x/\varepsilon-x^{2}/(2\varepsilon^2)  &\quad  if\ x\le\varepsilon
\end{array} \right.   
\end{equation*}
where $\varepsilon$ is chosen as $1/K$ in this paper. The SSMEL estimaotr $\hat{\theta}_{S}$ is calculated by minimizing the following objective function:
\begin{equation}\label{target_f}
\hat{\theta}_{S} = \arg\min\limits_{\theta\in\bar{\Theta}}f(\theta)
\end{equation}

We apply the two-layer coordinate decent algorithm in \citet{tang2014algo} to solve the problem. The inner layer of the algorithm is to find $\lambda$ by maximizing $f(\lambda,\theta)$ for a fixed $\theta$. The outer layer of the algorithm is to search for the optimal $\hat{\theta}_{S}$, and coordinate descent can be used to solve both layers.

The inner-layer involves maximizing $f(\lambda,\theta)$ as defined in Equation~\protect(\ref{inner_layer}) for a fixed $\theta$. Assuming the initial value of $\lambda$ is $\lambda^{(0)}$, we fix the other coordinates and calculate the value of $\lambda_j$, where $j=1,2,\cdots,r$ in the $(M+1)$th iteration, the 
$j$th component of $\lambda$ is given by

\begin{equation}\label{calc_lam}
\hat{\lambda}_{j}^{(M+1)}=\hat{\lambda}_{j}^{(M)}-\frac{\sum\limits_{k=1}^{K}\log_{*}^{'}\left(t_{k}^{(M)}\right)\cdot \bar{g}_{j}^{(k)}(\theta)}{\sum\limits_{k=1}^{K}\log_{*}^{''}\left(t_{k}^{(M)}\right)\cdot \left\{\bar{g}_{j}^{(k)}(\theta)\right\}^2}
\end{equation}
where $t_{k}^{(M)}=1+\bar{g}^{(k)}(\theta)^{T}\hat{\lambda}^{(M)}$, $\hat{\lambda}^{(M)}=(\hat{\lambda}_1^{(M)},\cdots,\hat{\lambda}_r^{(M)})^{T}$. The procedure is repeated with each of the $r$ elements of $lambda$ until convergence. At each step, it is crucial to optimize the objective function. If not, keep halving the step size until it is driving the objective function in the right direction. The procedure in Equation (\ref{calc_lam}) can be viewed as an optimization of a univariate sequence.

The outer layer can also be solved using a coordinate descent algorithm. At a given $\lambda$, the algorithm updates $\theta_t, t=1,\cdots,p$ by minimizing $f(\theta)$ defined in Equation~\protect(\ref{target_f}) with respect to $\theta_{t}$ with other $\theta_l$ is fixed, $l\neq t$. Assuming the initial value of $\theta$ is $\hat{\theta}^{(0)}$, the $(M+1)$th Newton update for $\theta_t$ is given by

\begin{equation}\label{calc_tet}
\hat{\theta}_{t}^{(M+1)}=
\hat{\theta}_{t}^{(M)}-\frac{\sum\limits_{k=1}^{K}\log_{*}^{'}\left(s_{k}^{(M)}\right)w_{kt}^{(M)}}{\sum\limits_{k=1}^{K}\left\{\log_{*}^{''}\left(s_{k}^{(M)}\right)\left(w_{kt}^{(M)}\right)^{2}+\log_{*}^{'}\left(s_{k}^{(M)}\right)z_{kt}^{(M)}\right\}}
\end{equation}
where $s_{k}^{(M)}=1+\lambda^{T}\bar{g}^{(k)}\left(\hat{\theta}^{(M)}\right)$, $w_{kt}^{(M)}=\lambda^T\partial\bar{g}^{(k)}\left(\hat{\theta}^{(M)}\right)/\partial\theta_{t}$, and $z_{kt}^{(M)}=\lambda^T\partial^2\bar{g}^{(k)}\left(\hat{\theta}^{(M)}\right)/\partial\theta_{t}^2$ with $\hat{\theta}^{(M)}=(\hat{\theta}^{(M)}_1,\cdots,\hat{\theta}^{(M)}_p)^{T}$. Note that Equation~(\ref{calc_tet}) $\lambda$ actually depends on $\hat{\theta}^{(M)}$ by definition~\protect(\ref{target_f}). This implies that upon updating one component $\theta_{t}$, $\lambda$ needs an update.  For the distributed data, we give the pseudo-code in Algorithm~\protect\ref{algo1}. The SSMEL is similarly simple to compute via Algorithm~\protect\ref{algo1} in the situation of large data on a single computer where subset information does not need to be sent between each device.
\begin{algorithm}
    \caption{The SSMEL for distributed data}\label{algo1}
    \renewcommand{\algorithmicrequire}{\textbf{Input:}}
    \renewcommand{\algorithmicensure}{\textbf{Output:}}
    \begin{algorithmic}[1]
		\STATE Set the iteration counter $M=0$, and initialize  $\theta^{(0)}$ and $\lambda^{(0)}$, threshold $\gamma=10^{-4}$
		\REPEAT
		\STATE  Each local device evaluates $\bar{g}^{(k)}\left(\hat{\theta}^{(M)}\right),w_{k}^{(M)},z_{k}^{(M)}$ and sends to the central processor
              \FOR{$t=1\ to\ p$}
                 \STATE (1) Calculate $\hat{\theta}_{t}^{(M+1)}$ as in Equation~\protect($\ref{calc_tet}$)
                 \STATE (2) Update all $\lambda_{j}$ as in Equation~\protect($\ref{calc_lam}$) for $j=1,\cdots,r$ coordinate-wise
              \ENDFOR
		\STATE The central processor sends $\hat{\theta}^{(M+1)}$ to the  local machines
             \STATE M $\leftarrow$ M+1
		\UNTIL $\max\limits_{1\le t\le p}\left(\left|\theta_{t}^{(M+1)}-\theta_{t}^{(M)}\right|\right)<\gamma$
		\ENSURE  $\hat{\theta}^{(M+1)}$
	\end{algorithmic}
\end{algorithm}

\begin{rmk}
It can be seen that the one-shot approach needs to perform numerous optimization operations in parallel, but the SSMEL just requires optimizing a single objective function. Therefore, our approach is convenient and efficient if massive data can be loaded into memory and processed on a single computer, and it is appropriate for generic computing systems.
\end{rmk}

\begin{rmk}
Algorithm \ref{algo1} is a simple implementation of the SSMEL applied to distributed data, it may be thought of as an empirical likelihood in the context of the iterative approach, and, in addition to being able to ensure good estimation efficiency, the SSMEL has one major advantage over previous approaches: easy and powerful statistical inference. It is worth noting that the efficiency of the SSMEL is not limited by the number of devices. The data on a single device may be randomly divided into several datasets if the number of devices is too little, which implies that in the distributed scenario $K$ may not be equal to the actual number of devices.
\end{rmk}

\section{Simulations}
We show how the SSMEL approach performs through several simulations in this section. The SSMEL's behavior in reducing computation time for large data sets is examined in the first example, which also compares the estimation accuracy and computation times of the SSMEL, DEL, and SSEL in various situations. In the second illustration, three different splitting numbers illustrate the estimation accuracy of the SSMEL, DEL, and SSEL with altering parameter dimension $p$. The final example offers the findings of the SSMEL, DEL, and SSEL hypothesis testing. In these instances, the centralized empirical likelihood (CEL) represents the empirical likelihood for the entire dataset. Owing to \citet{ssel2020} does not explicitly explain how to optimize the components of the SSEL function in parallel, and the highlights in \citet{zhou2022distributed} are similar to \citet{ssel2020}, we use the algorithm in \citet{zhou2022distributed} to implement the SSEL. All simulations were implemented in R, and parallel computing using \texttt{parallel} package, which is included in R.

\subsection{Example 1: estimating the parameters of normal distribution}
The data $X_1,\cdots,X_n$ are produced from a normal distribution $N(\mu, \sigma^2)$, where the unknown parameters $\mu$ and $\sigma$ are generated at random from the uniform distributions $Unif(-2,2)$ and $Unif(0.5,2)$, respectively. The random variable $X$ satisfies the following moment conditions:
\begin{equation*}
 \mathbb{E}\big[g(X,\theta_{0})\big]=\mathbb{E}
\begin{pmatrix}
 \mu - X \\
\sigma^{2}-(X-\mu)^{2} \\
X^{3}-\mu(\mu^{2}+3\sigma^{2})
\end{pmatrix}
=0.
\end{equation*}

\begin{case}
To assess the efficacy of the proposed approach in terms of reducing computation time for massive data, we choose the full dataset size at $n=200000$ and vary $K=[10,50,100,500,1000, 5000]$, with $K=200000$ being equivalent to the CEL. This procedure with 500 replications, the mean square error (MSE) for $\mu$ and $\sigma$, and total computation time (TCT) are recorded in Table~\protect\ref{e1c1}.

\begin{table}[]
\centering
\caption{MSE$(\times10^{-6})$ of different estimators and total CT under varying $K$}\label{e1c1}
\begin{tabular}{cccc}
   \toprule
    $K$ &MSE of $\mu$ &MSE of  $\sigma$ & TCT(s)\\
   \midrule
        10    &11.304(16.832) &5.350(10.058)&240.8307 \\
        50    &9.304(14.358)  &4.876(9.325)  &196.4065 \\
       100   &9.008(14.093)  & 4.732(9.176) &196.6988  \\
       500   &9.095(14.506)  &4.656(8.842)  &307.2509 \\
      1000  &9.130(14.592)  &4.679(8.970)  &457.3839\\
      5000  &9.124(14.481)  &4.677(9.017)  &1826.269\\
    200000&9.066(14.378)  &4.695(9.043)  &48792.91 \\
   \bottomrule
\end{tabular}
\end{table}

Table~\protect\ref{e1c1} shows that, with the proper $K$, our proposed approach can effectively reduce the computation time and achieve the estimation accuracy under the full dataset. This confirms the conclusion of Corollary \ref{crl1}. A smaller value of $K$ is not recommended, as this would over-compress the sample information, and cause poor estimation accuracy and non-optimal computational efficiency.
\end{case}

\begin{case}
In this case, the performance of the CEL, SSMEL, DEL, and SSEL is compared by evaluating their accuracy for four different settings with 1000 repetitions. To further explore the sensitivity of each method concerning $K$ and $m$, we did not consider setting $n$ extremely large, and the subsequent simulation settings were similarly based on this consideration. The logarithmic mean squared error (log-MSE) is shown in Fig.~\protect\ref{compare_method}.
\begin{enumerate}
\item[(a)] Fixing the subset size $m$, the split size $K$ increases with the size of the full dataset $n$. We consider $m$ = [50,100] and vary $n$ = [1000,2000,3000,4000,5000].
\item[(b)] Fixing the split size $K$, the subset size $m$ increases with the size of full dataset $n$. We consider $K$ = [10,50] and vary $n$ = [1000,2000,3000,4000,5000].
\item[(c)] Fixing the size of full dataset at $n=12000$ and vary $K$ = [10,20,40,80,100,120].
\item[(d)]  Fixing the size of full dataset at $n=12000$ and vary $m$ = [100,200,400,800,1000,1200].
\end{enumerate}

\begin{figure}[htbp]
    \begin{minipage}[t]{0.33\linewidth}
        \centering
        \includegraphics[width=0.95\textwidth]
        {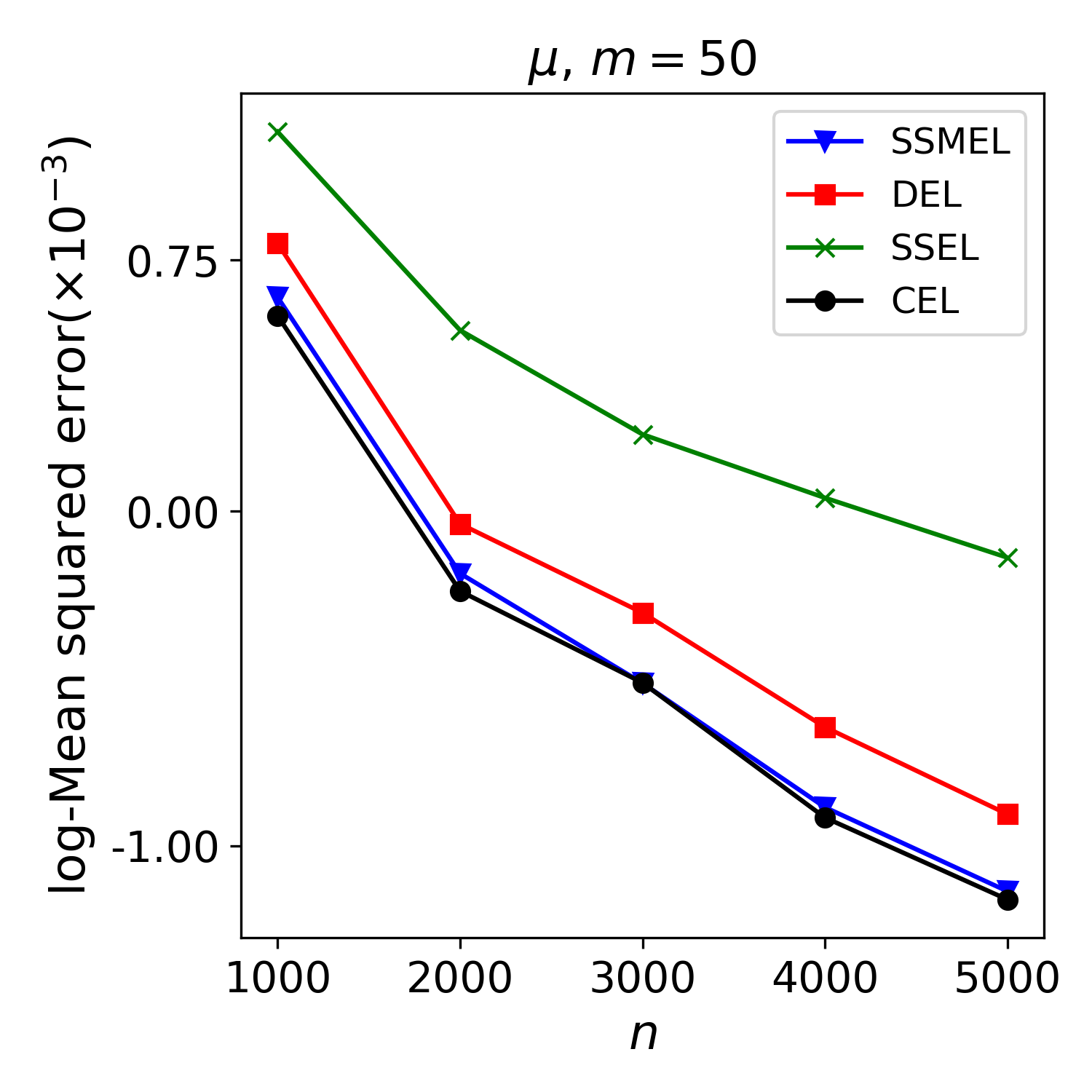}
    \end{minipage}%
    \begin{minipage}[t]{0.33\linewidth}
        \centering
        \includegraphics[width=0.95\textwidth]
        {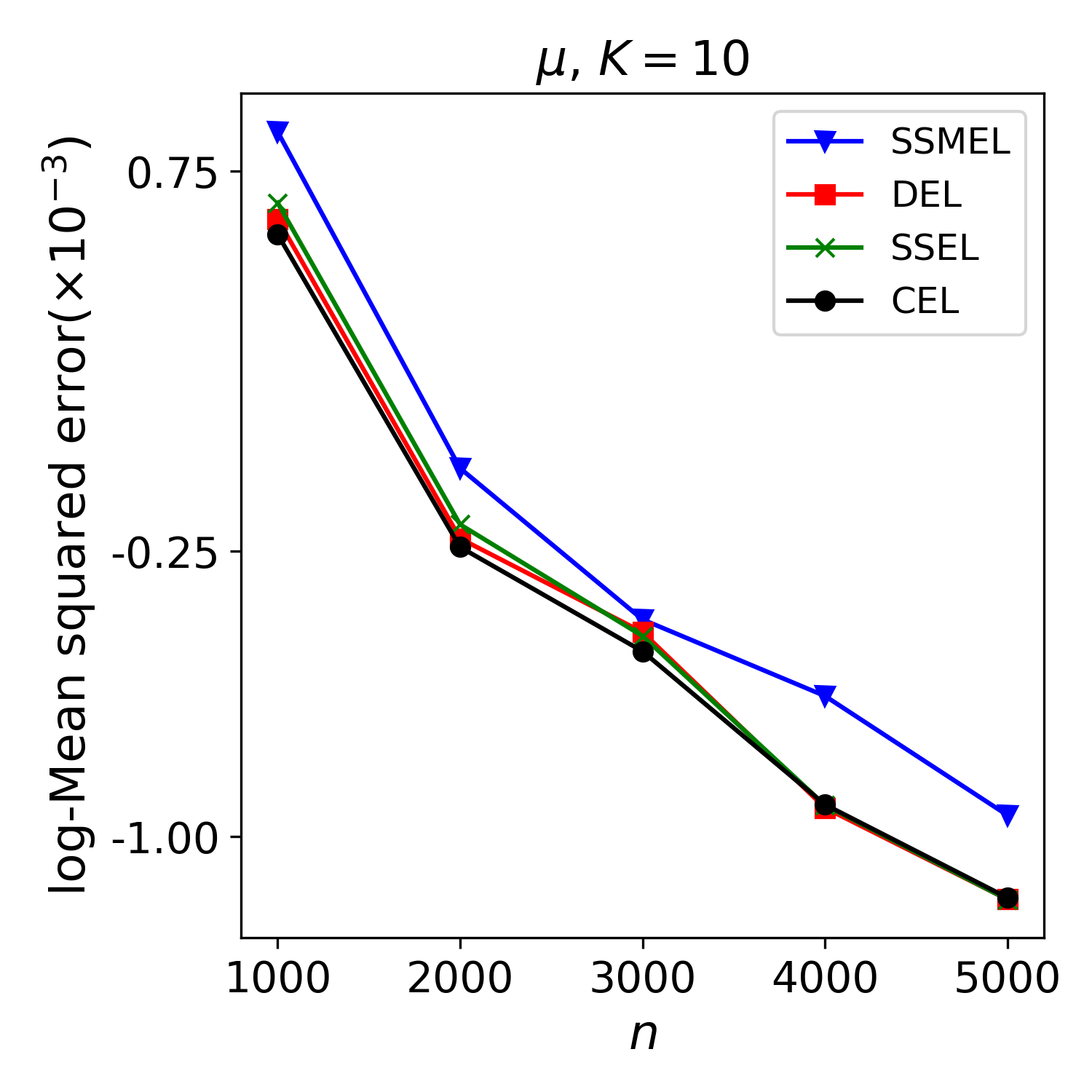}
    \end{minipage}%
    \begin{minipage}[t]{0.33\linewidth}
        \centering
        \includegraphics[width=0.95\textwidth]
        {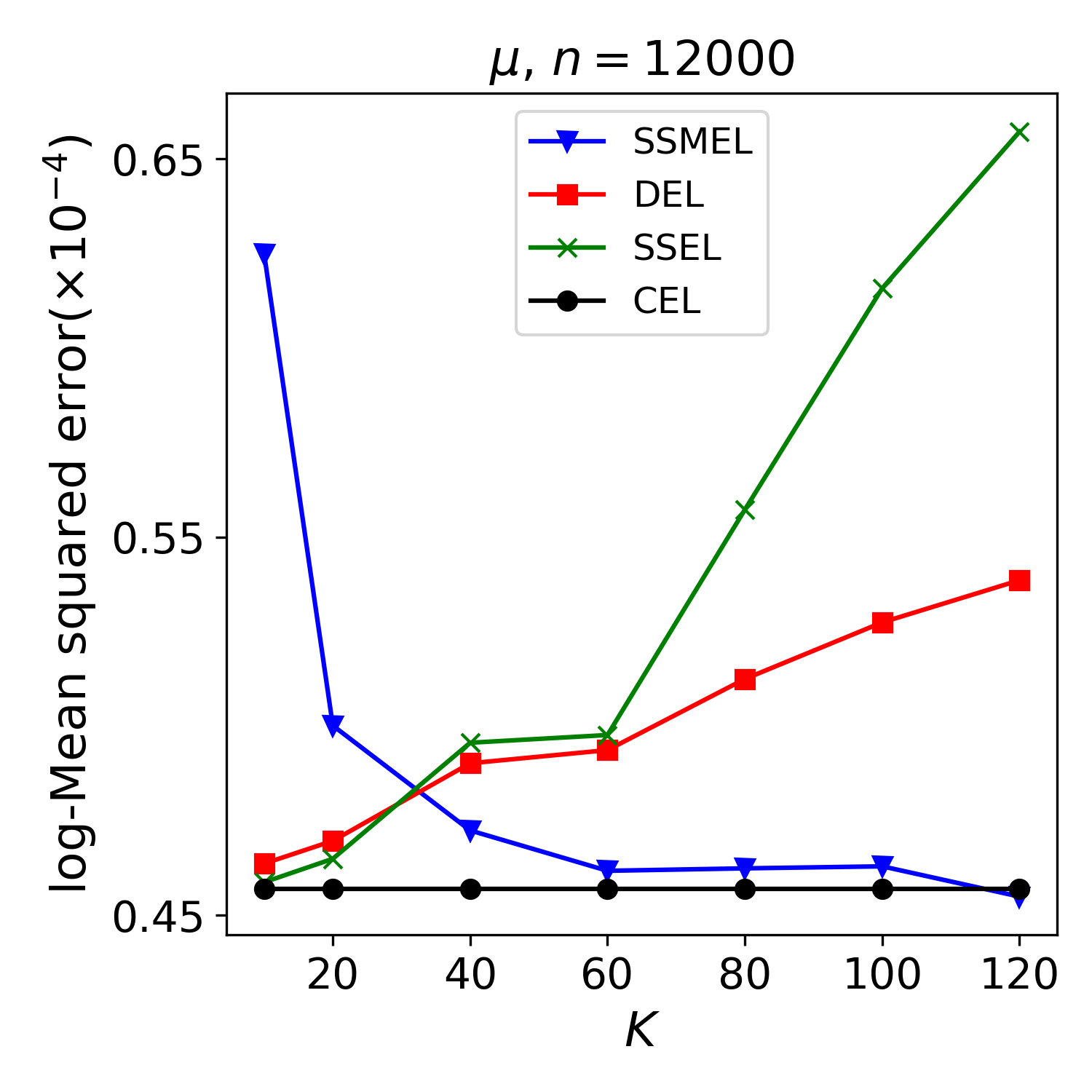}
    \end{minipage}%
   
     \begin{minipage}[t]{0.33\linewidth}
        \centering
        \includegraphics[width=0.95\textwidth]
        {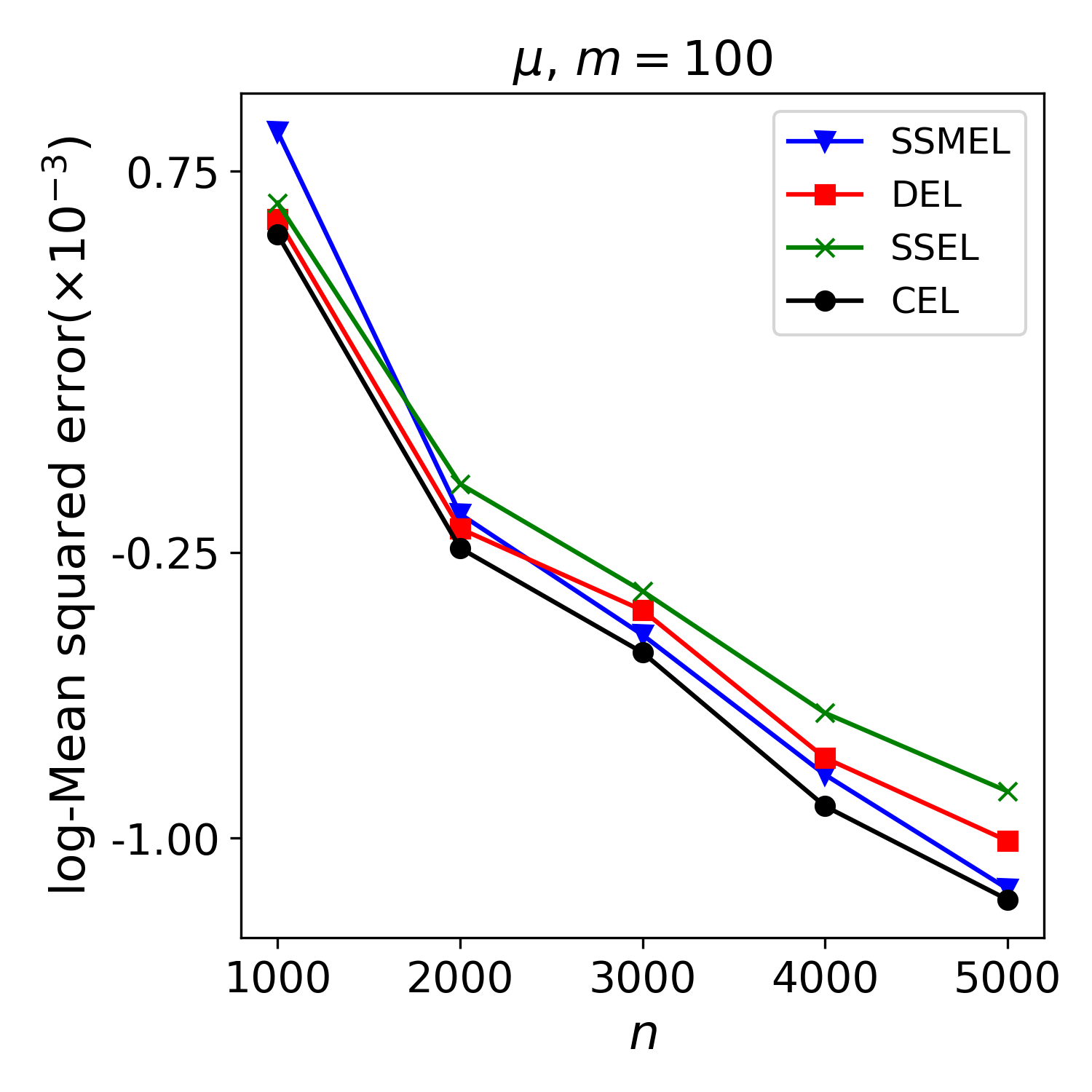}
    \end{minipage}%
    \begin{minipage}[t]{0.33\linewidth}
        \centering
        \includegraphics[width=0.95\textwidth]
        {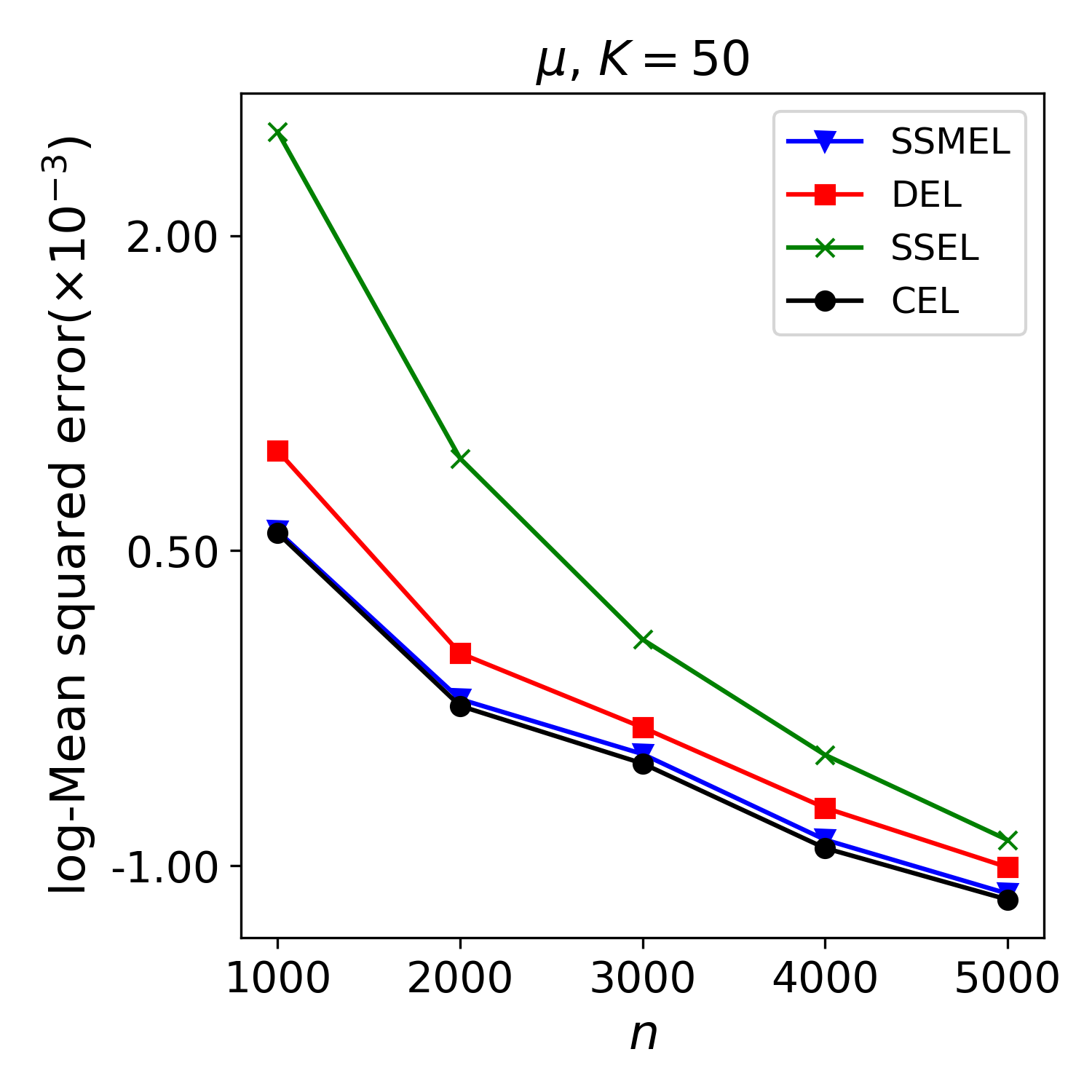}
    \end{minipage}%
        \begin{minipage}[t]{0.33\linewidth}
        \centering
        \includegraphics[width=0.95\textwidth]
        {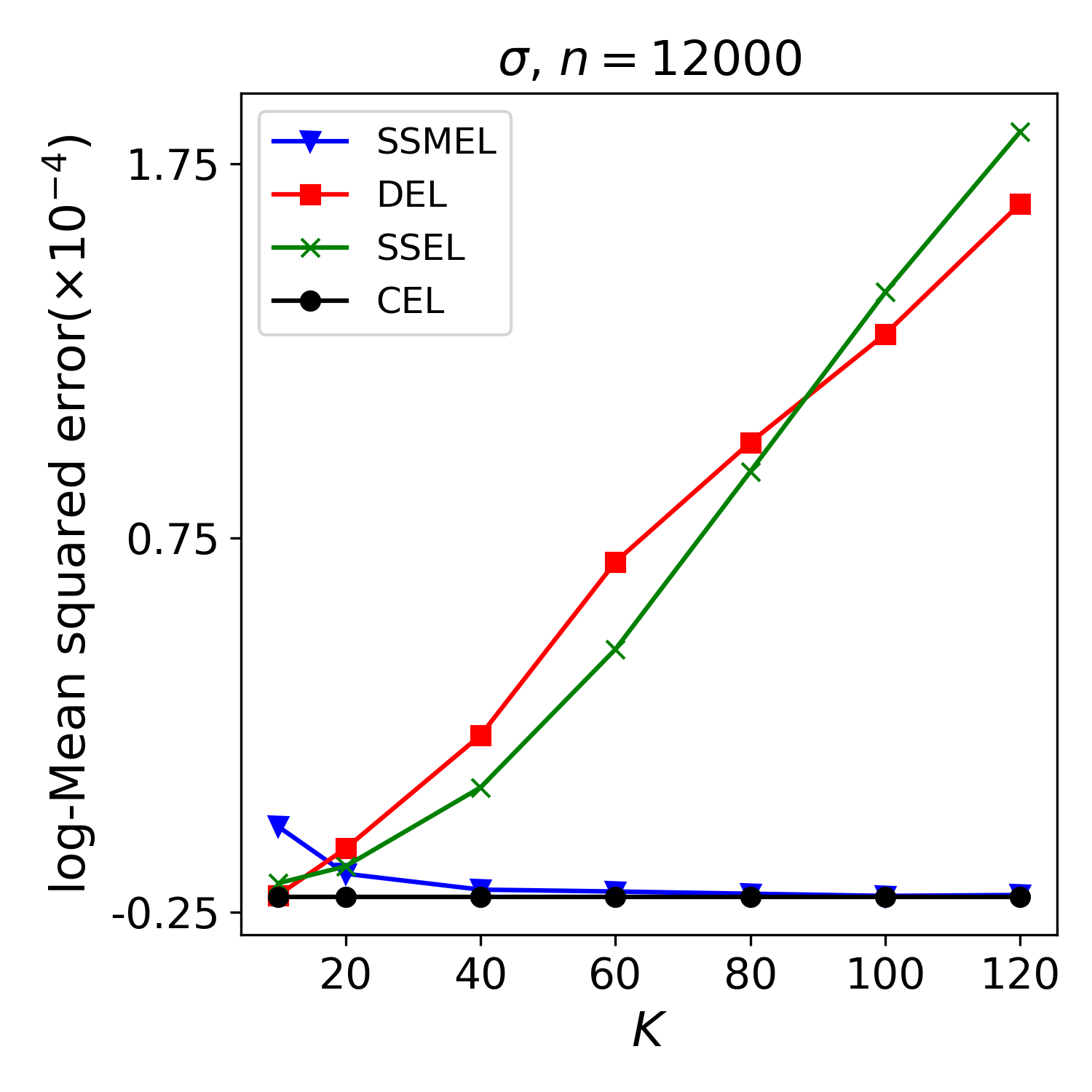}
    \end{minipage}%
 
     \begin{minipage}[t]{0.33\linewidth}
        \centering
        \includegraphics[width=0.95\textwidth]
        {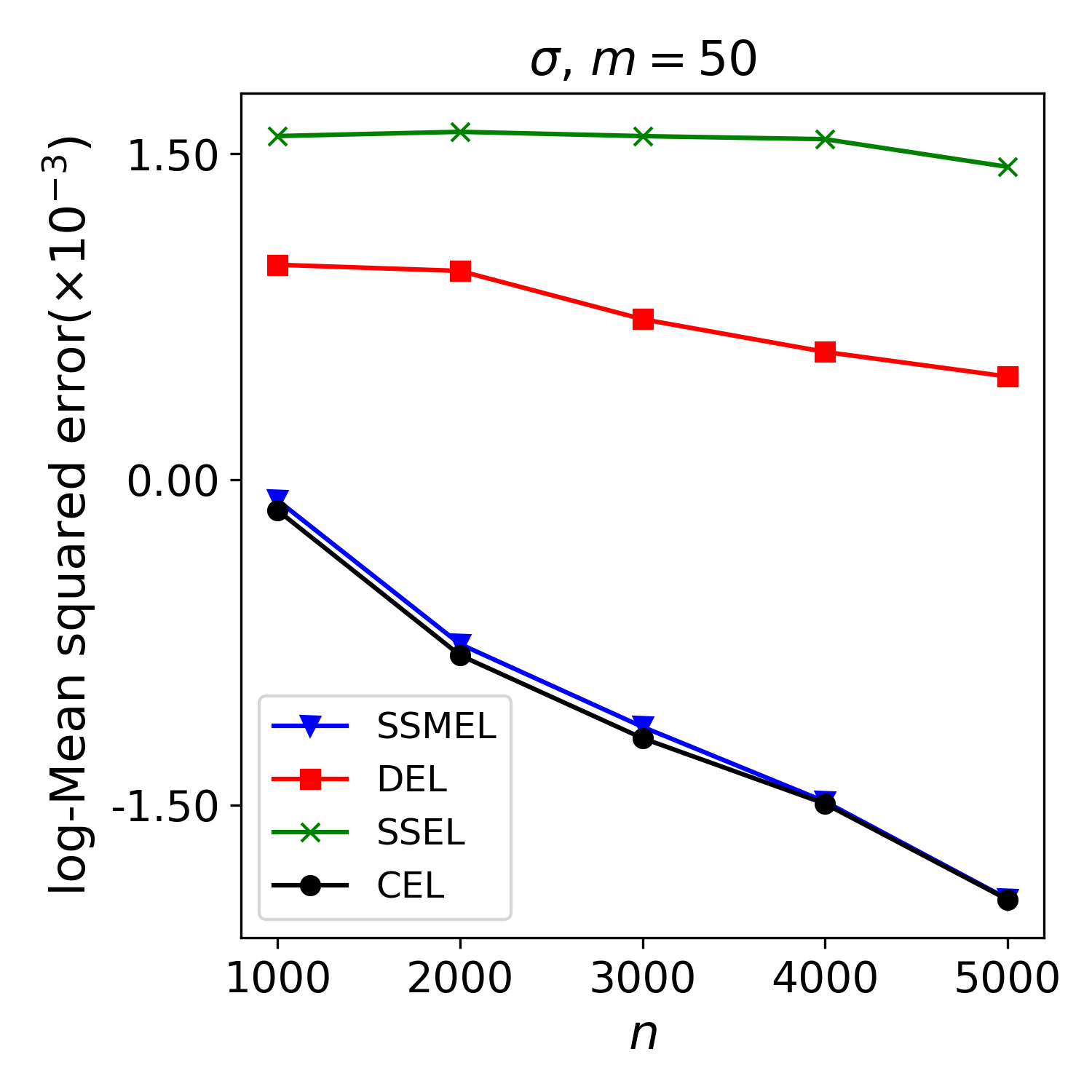}
    \end{minipage}%
    \begin{minipage}[t]{0.33\linewidth}
        \centering
        \includegraphics[width=0.95\textwidth]
        {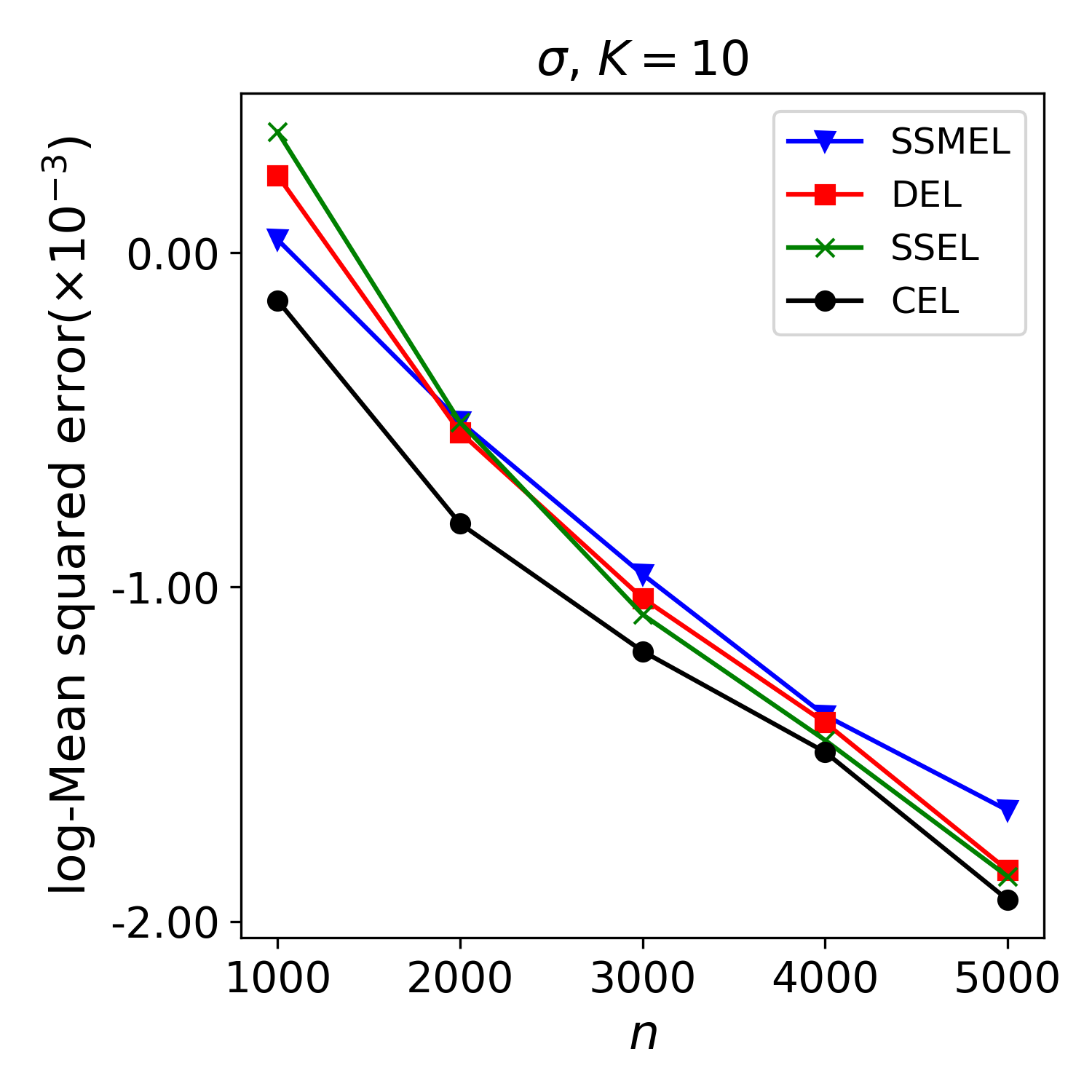}
    \end{minipage}%
	\begin{minipage}[t]{0.33\linewidth}
        \centering
        \includegraphics[width=0.95\textwidth]
        {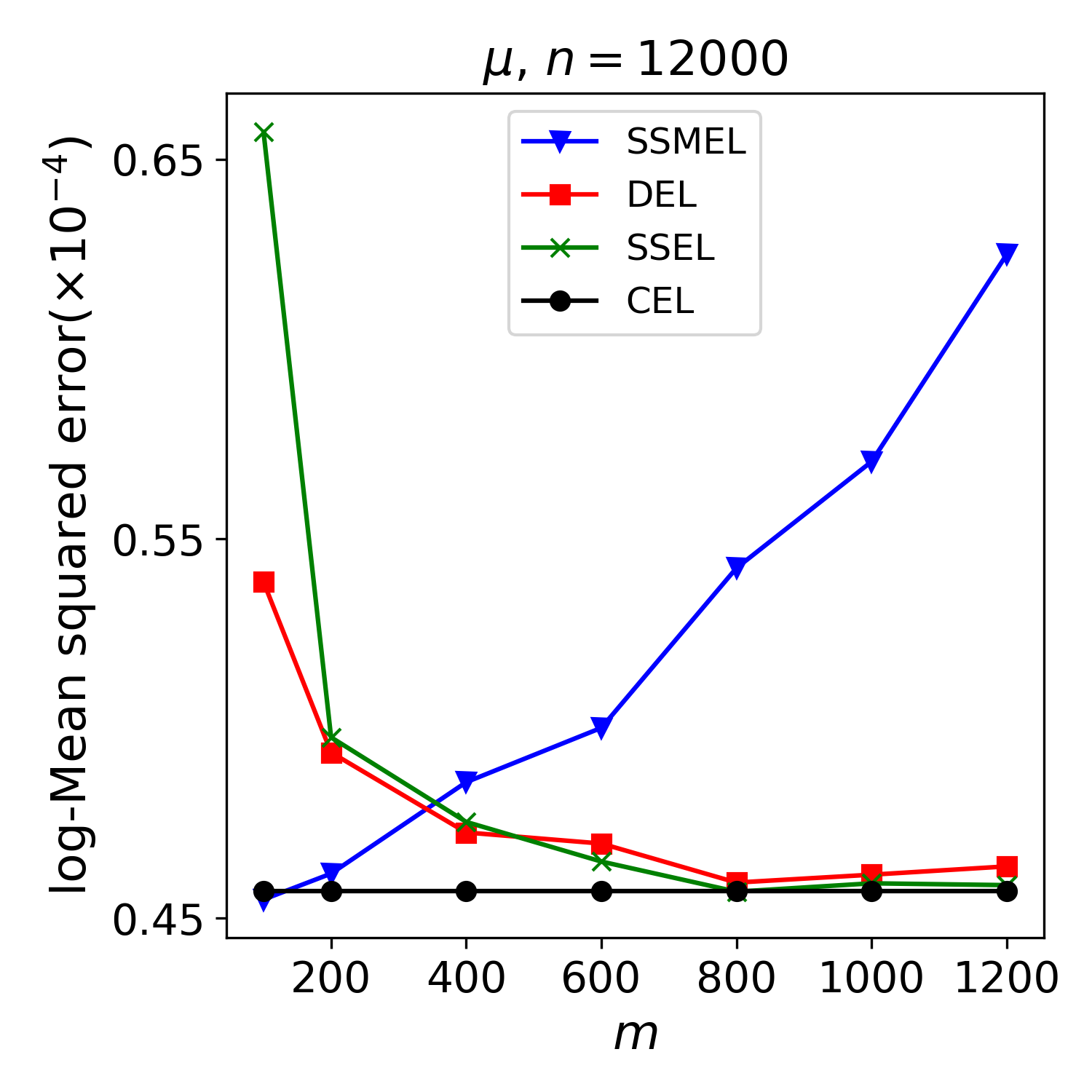}
    \end{minipage}%
    
      \begin{minipage}[t]{0.33\linewidth}
        \centering
        \includegraphics[width=\textwidth]
        {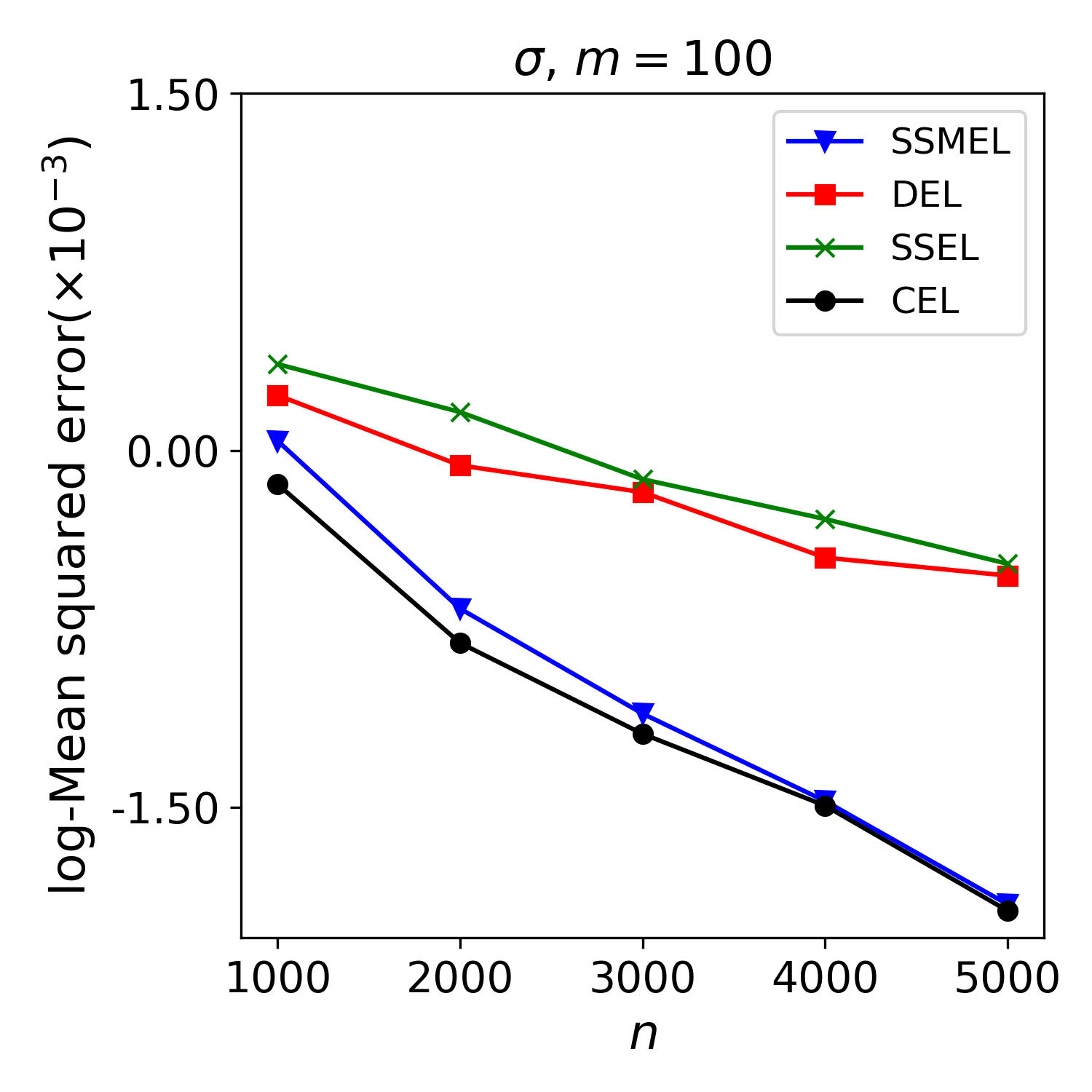}
    \end{minipage}%
    \begin{minipage}[t]{0.33\linewidth}
        \centering
        \includegraphics[width=\textwidth]
        {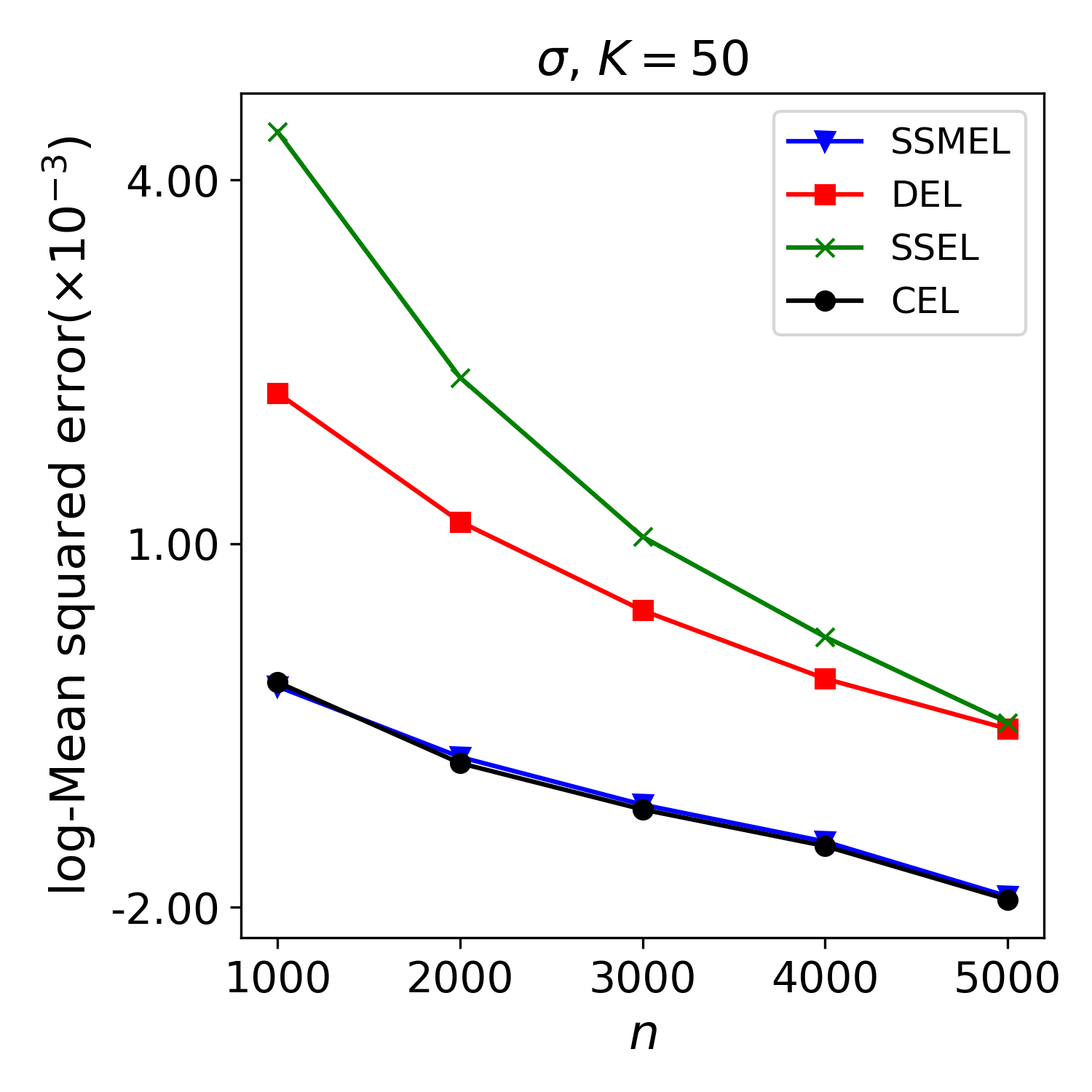}
    \end{minipage}%
	\begin{minipage}[t]{0.33\linewidth}
        \centering
        \includegraphics[width=\textwidth]
        {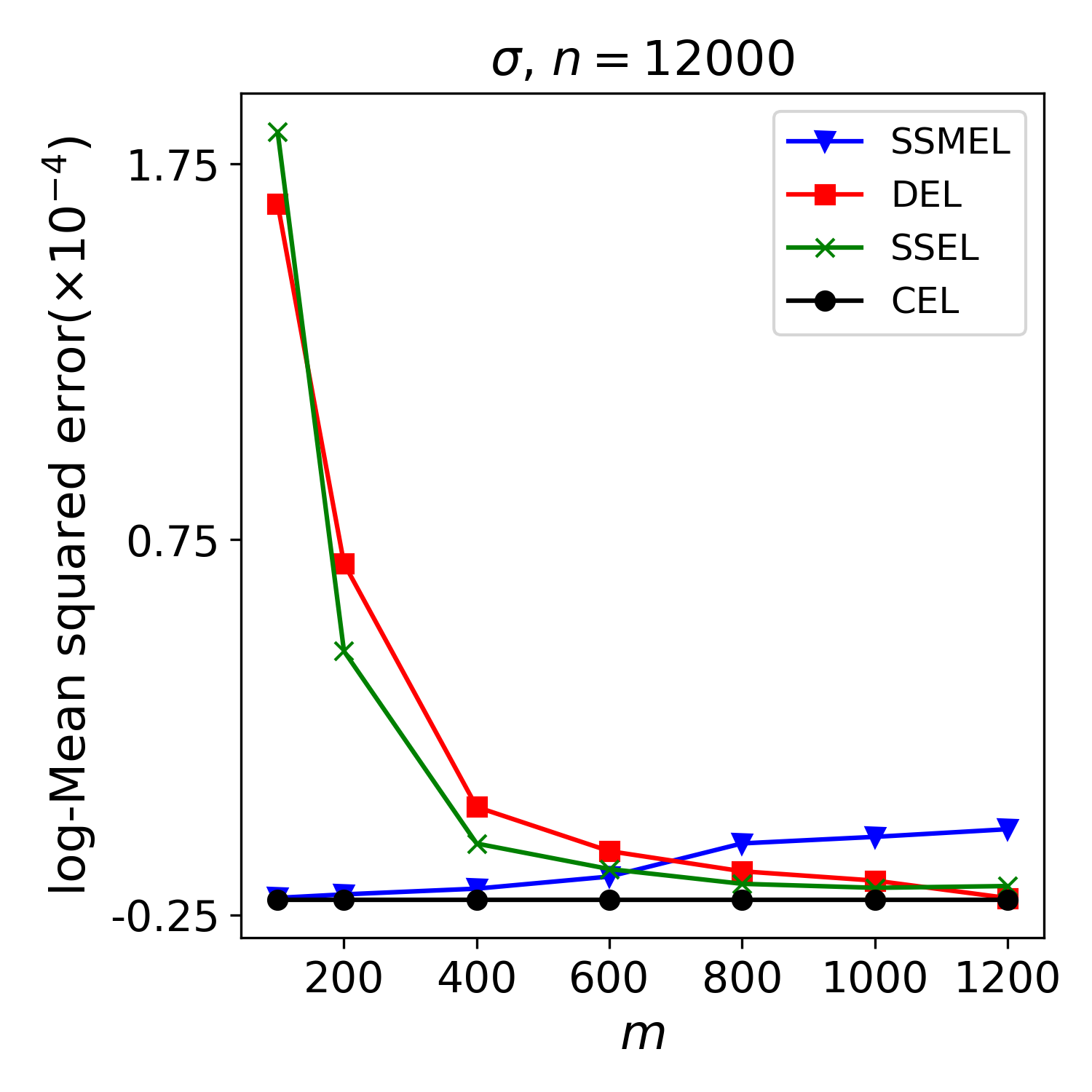}
    \end{minipage}%
    \caption{Plots of log-MSE with different settings. The first column represents the log-MSE of setting $(a)$ at different $m$, and the second column represents the log-MSE of setting $(b)$ at different $K$. In the third column, rows 1-2 represent the log-MSE for setting $(c)$ and rows 3-4 for setting $(d)$. }\label{compare_method}
\end{figure}

From the first and second columns of Fig.~\protect\ref{compare_method}, we can see that for fixed $K$ and $m$, the log-MSE of the SSMEL on $\mu$ and $\sigma$ are closer to the behavior of the CEL when the total sample increases. The SSEL and DEL perform poorly for the nonlinear statistic $\sigma$. With a fixed $n$, it can be seen from the third column of Fig.~\protect\ref{compare_method} that when $K$ increases to 100, the SSMEL achieves the estimation efficiency under the full dataset. When $m$ is large i.e. $K$ is small, the log-MSE of the SSMEL tends to increase due to excessive compression of information. Overall, the SSMEL performs more robustly compared to other methods.
\end{case}

\begin{case}
In this case, we compare the performance of each method in terms of the reduction of computation time. We consider $n=12000$ and 500 repetitions, and the results are presented in Table~\protect\ref{compare_time}. The SSMEL decreases computation time more effectively than the DEL and SSEL, as shown by Table~\protect\ref{compare_time}. The SSMEL displays superior computational efficiency without relying on parallel computing hardware, which is worth mentioning. The DEL and SSEL, on the other hand, might use parallel processing to split up the computation time, but as $K$ rises, their estimation accuracy might suffer.

\begin{table}[]
\centering
\caption{The MSE and computation time for different methods at varying $K$}\label{compare_time}
\resizebox{\textwidth}{!}{
\begin{tabular}{ccccccccc}
\toprule
Method &  & $K$    &  & MSE($\times10^{-4}$) of $\mu$& &MSE($\times10^{-5}$) of $\sigma$ &   & CT(s) \\
\midrule
CEL       &  & -----  &  &1.3254(2.2920) &  &8.5468(14.0489)  &    &3.4857(8.0277)\\ 
DEL       &  & 10    &   &1.3193(2.2609)&  &8.9280(14.8133)  &   &2.1655(2.4535)\\
            &  & 50     &   &1.3318(2.2862)&  &16.6670(38.1688)&   &1.4172(1.2647)\\
            &  & 100   &  &1.3729(2.4160)&   &38.2400(47.1748)&   &1.1681(0.9020)\\  
SSEL    &  & 10    &   &1.3206(2.2613)&   &8.7211(14.5149)  &   &3.6237(3.6249)\\
           &  & 50    &   &1.3614(2.3521) &   &13.4378(21.5337) &   &1.7358(1.7520)\\  
            &  & 100  &  &1.5293(2.6973) &   &39.4634(47.6825) &   &1.3428(1.2913)\\   
SSMEL  &  & 10    &  &1.6680(2.6744) &   &11.6354(29.1214) &   &0.1421(0.1055)\\ 
            &  & 50   &  &1.3242(2.2625) &   &8.7477(14.0602)   &   &0.1586(0.1883)\\
            &  & 100  &  &1.3142(2.2660) &  &8.6603(14.1200)    &   &0.1682(0.1760)\\
\bottomrule
\end{tabular}
}
\end{table}

\end{case}

\subsection{Example 2: regression models}
In this example, we take into account estimating the coefficients of a linear regression model with various parameter numbers.,
\begin{equation}
Y_{i} =Z_{i}^{T}\beta+\varepsilon_{i},\quad i=1,\cdots,n
\end{equation}
where $\beta=(\beta_0,\beta_1,\cdots,\beta_p)^{T},\varepsilon_{i}$ and $Z_{i}=(1,X_{i1},\cdots,X_{ip})^{T}$ are independent, $\varepsilon\stackrel{\text{i.i.d}}{\sim}N(0,1)$, $X_{i}=(1,X_{i1},\cdots,X_{ip})^{T}\stackrel{\text{i.i.d}}{\sim}N(0,\Sigma_{p})$, where $\Sigma_{p}$ is a $p\times p$ matrix with main diagonal being 1 and off-diagonal being $\rho$.
we refer to the setting in \citet{del2022} that $p=[4,8,18]$, $\beta_{0}=(1,5,4,3,2,\textbf{1}_{p-4}^{T})^{T}$ with $\textbf{1}_{p-4}^{T}=(1,\cdots,1)^{T}$ for $p>4$, and $\beta_{0}=(1,5,4,3,2)$ for $p=4$. $\rho=[0,0.2,0.5,0.8]$ in $\Sigma_{p}$. We fix the full dataset size at $n=20000$ and $K=[10,50,100]$. We compare the performance of SSMEL, DEL, and SSEL estimators in terms of empirical MSE, i.e. the mean of $\Vert \hat{\beta} - \beta_0 \Vert^2$. Table~\protect\ref{reg_table} summarizes these results based on 500 replications. From Table~\protect\ref{reg_table}, we can see that the MSE of the SSMEL is very close to that of the CEL in most cases. The SSMEL exhibits poor performance when $K$ is close to $p$ because the approach formally uses only $K$ samples. When $K$ is significantly greater than $p$, the SSMEL is computationally simple and the estimation is robust compared to other methods in various cases.
\begin{table}[]
\centering
\caption{The MSE($\times10^{-4}$) of the SSMEL, DEL, SSEL, and CEL estimators of $\beta$}\label{reg_table}
\resizebox{\textwidth}{!}{
\begin{tabular}{cccccccccccccccc}
\toprule
Method             &     & \multicolumn{3}{c}{SSMEL} &  & \multicolumn{3}{c}{DEL} &  & \multicolumn{3}{c}{SSEL} & &  CEL \\ \cline{1-1} \cline{3-5} \cline{7-9} \cline{11-13} \cline{15-15}
\rule{0pt}{16pt}$p$                 & $\rho$   & $K=10$     & $K=50$     & $K=100$    &  & $K=10$     & $K=50$    & $K=100$    &  & $K=10$     & $K=50$     & $K=100$ & & ------ \\ \midrule
\multirow{4}{*}{4} 
                   & 0   &2.470&2.470&2.470&  &2.472&2.507&2.548&  &2.470&2.570&2.693&  &2.470\\
                   & 0.2 &2.794&2.729&2.729&  &2.738&2.777&2.810&  &2.753&2.862&2.890& &2.729\\
                   & 0.5 &3.661&3.653&3.653&  &3.669&3.678&3.696&  &3.680&3.807&8.658& &3.653\\
                   & 0.8 &8.117&8.125&8.125&  &8.133&8.310&8.416&  &8.120&8.583&8.981& &8.125\\ 
\multirow{4}{*}{8} 
                    & 0   &7.345&4.608&4.608&  &4.644&4.784&4.922&  &4.674&4.882&5.105& &4.608\\
                   & 0.2 &10.554&5.278&5.278&  &5.329&5.462&5.608&  &5.318&5.519&5.796& &5.278\\
                   & 0.5 &13.674&7.602&7.602&  &7.676&7.865&8.068&  &7.722&8.174&8.546& &7.602\\
                   & 0.8 &28.561&18.715&18.715&  &18.883&19.238&19.876&
 &18.951&20.112&21.097& &18.715\\                    
 \multirow{4}{*}{18} 
                    & 0   &   ------     &9.643&9.644&  &9.742&10.020&10.611&  &9.883&10.695&11.925&  &9.644\\
                   & 0.2 &    ------   &11.655 &11.655&  &11.717&12.080&12.770&  &11.887&12.917&14.260&  &11.655\\
                 & 0.5 &    ------  &17.893&17.893&  &18.044&18.604&19.725&  &18.328&19.888&22.075& &17.893\\
                   & 0.8 &   ------     &43.440&43.424&  &44.212&45.719&48.621&  &44.405&48.310&54.284& &43.424\\               
\bottomrule
\end{tabular}
}
\end{table}

\begin{rmk}
Since a plane can only be formed by three points in two dimensions, the parameter space of the SSMEL is the convex hull of $\{\bar{g}^{(k)}(\theta), k=1,\cdots, K\}$. Therefore, when $K$ is close to $p$, the convex hull formed by $K$ points is unable to effectively encompass $p$-dimensional vectors, resulting in poor estimation performance.
\end{rmk}

\subsection{Example 3: binary normal distribution hypothesis test}
In this example, we examine the performance of the hypothesis test for the SSMEL. We generate data $(X,Y)$ from the bivariate normal distribution $N(\mu_1,\sigma_1^2,\mu_2,\sigma_2^2,\rho)$. We use a dataset size of 10000 and 500 replications with parameter values $\mu_1=\mu_2=0$, $\sigma_1=\sigma_2=1$, and $\rho=0.5$. The random vector $(X,Y)$ satisfies the following moment conditions:
 \begin{equation*}
  \mathbb{E}\big[g(X,Y;\theta_{0})\big]=\mathbb{E}
\begin{pmatrix}
 \mu_1 - X \\
\mu_2 - Y \\
\sigma_1^{2}-(X-\mu_1)^{2} \\
\sigma_2^{2}-(Y-\mu_2)^{2} \\
(X-\mu_1)(Y-\mu_2)-\rho\sigma_1\sigma_2
\end{pmatrix}.    
 \end{equation*}

The hypothesis testing of the DEL is challenging due to the failure to meet Wilks' theorem. To address this problem, \citet{ma2022stat} recently presented a statistical inference method for the one-shot estimator of average aggregation. The primary objective is to test the mean of the estimators of each subset using empirical likelihood. We applied the method to the DEL for hypothesis testing and compared it with the SSMEL, SSEL, and CEL. We consider the following null hypothesis:
\begin{itemize}
\item$H_{01}:\theta = (\mu_1,\mu_2,\sigma_1,\sigma_2,\rho) = (0,0,1,1,0.5)$
\item$H_{02}:\psi = (\mu_1,\sigma_1) = (0,1)$
\item$H_{03}:\rho=0.5$
\end{itemize}

We choose $K=[5,10,20,40,80,100]$, the false rejection rate at nominal levels $\alpha=0.05$ for each method are recorded in Table~\protect\ref{alpha005} and  the empirical frequencies of $\rho\notin\{\rho:\mathcal{W}(\rho)=2(\ell_{S}(\rho)-\ell_{S}(\hat{\rho}))\le\chi^{2}_{1,0.95}\}$ for a sequence of $\rho$ values in Table~\protect\ref{power_analy}. From Table~\protect\ref{alpha005}, we can see that the false rejection rate when using the SSMEL is also affected by the choice of $K$. The false rejection rate of the SSMEL closes the nominal level and the results of the full dataset as $K$ increase, and the change becomes subtle when $K$ grows to a certain level ($K=100$). In contrast to the SSMEL, the SSEL is negatively impacted by larger values of $K$, as the false rejection rate deviates further away from the nominal level. The DEL is a more complex method compared to others, as it involves a trade-off between the one-shot estimation accuracy and the error of the asymptotic $\chi^{2}$ distribution of the empirical log-likelihood function. From Table~\protect\ref{power_analy} we can see that the test power of the SSMEL has a consistent performance with the empirical likelihood under the full datasets. Therefore, the SSMEL is an effective alternative to the CEL to address statistical inference challenges with massive data.

\begin{table}[]
\centering
\caption{Proportion of false rejections at $\alpha=0.05$ by varying $K$}\label{alpha005}
\resizebox{\textwidth}{!}{
\begin{tabular}{cccccccccccccccc}
\toprule
\multirow{2}{*}{$K$} &  & \multicolumn{4}{c}{$H_{01}$}                 &  & \multicolumn{4}{c}{$H_{02}$}                 &  & \multicolumn{4}{c}{$H_{03}$}                 \\ \cline{3-6} \cline{8-11} \cline{13-16} 
\rule{0pt}{16pt}                  &  & CEL               & SSMEL & DEL & SSEL &  & CEL               & SSMEL & DEL & SSEL &  & CEL               & SSMEL & DEL & SSEL \\
\midrule                   
5                  &  & \multirow{7}{*}{0.050} &------&------&0.050&  & \multirow{7}{*}{0.054} &------
&------ &0.048&  & \multirow{7}{*}{0.058} &------&------&0.060\\
10                 &  &                   &0.676&0.674&0.052&  &                   &0.198&0.208&0.050&  &                   &0.100&0.124&0.062\\
20                 &  &                   &0.266&0.272 &0.062&  &                   &0.096&0.108&0.052&  &                   &0.068&0.080&0.056\\
40                 &  &                   &0.122&0.146&0.090&  &                   &0.054&0.096&0.060&  &                   &0.052&0.070&0.064\\
80                 &  &                   &0.052&0.182&0.284&  &                   &0.054&0.146&0.186&  &                   &0.062&0.062&0.066\\
100                &  &                   &0.052&0.674&0.502&  &                   &0.058&0.208&0.334&  &                   &0.054&0.088&0.088\\
 \bottomrule
\end{tabular}
}
\end{table}

\begin{table}[]
\centering
\caption{The empirical frequency of different methods that a given value of $\rho$ does not fall in the 95\% confidence set. The truth is $\rho=0.5$.}\label{power_analy}
\resizebox{\textwidth}{!}{
\begin{tabular}{ccccccccccccccccccccc}
\toprule
Method &  & $K$     &  & 0.46 &  & 0.47 &  & 0.48 &  & 0.49 &  & 0.50 &  & 0.51 &  & 0.52 &  & 0.53 &  & 0.54 \\
\midrule
CEL    &  & ----- &  &1.000&  &0.978&  &0.756&  &0.246&  &0.058&  &0.266&  &0.784&  &0.978&  &1.000\\

 DEL   &  & 10    &  &1.000&  &0.974&  &0.786&  &0.330&  &0.100&  &0.342&  &0.784&  &0.964&  &1.000\\
         &  & 50    &  &1.000&  &0.982&  &0.778&  &0.294&  &0.058&  &0.264&  &0.724&  &0.966&  &1.000\\
       &  & 100   &  &1.000&  &0.982&  &0.784&  &0.336&  &0.074&  &0.222&  &0.670&  &0.954&  &0.996\\

SSEL    &  & 10    &  &1.000&  &0.978&  &0.746&  &0.252&  &0.058&  &0.278&  &0.790&  &0.972&  &1.000\\
         &  & 50    &  &1.000&  &0.978&  &0.752&  &0.288&  &0.071&  &0.308&  &0.770&  &0.980&  &1.000\\
       &  & 100   &  &1.000&  &0.978&  &0.760&  &0.308&  &0.088&  &0.340&  &0.796&  &0.980&  &0.998\\

SSMEL &  & 10    &  &1.000&  &0.968&  &0.780&  &0.318&  &0.100&  &0.348&  &0.796&  &0.970&  &1.000\\
          &  & 50    &  &1.000&  &0.982&  &0.746&  &0.244&  &0.058 &  &0.286&  &0.770&  &0.972&  &1.000\\
       &  & 100   &  &1.000&  &0.982&  &0.736&  &0.244&  &0.054&  &0.284&  &0.788&  &0.982&  &1.000\\
\bottomrule
\end{tabular}
}
\end{table}

\section{Real data analysis}
\subsection{Protein dataset}
Physicochemical properties of protein tertiary structure dataset\footnote{\url{https://archive.ics.uci.edu/ml/datasets/Physicochemical+Properties+of+Protein+Tertiary+Structure}} are taken from CASP 5-9. There are 45730 decoys and sizes varying from 0 to 21 Armstrong, which aims to predict the size of the residue (RSMD). The explanatory variables are as follows: $X_1$, total surface area; $X_2$, non-polar exposed area; $X_3$, fractional area of exposed non-polar residue; $X_4$, fractional area of the exposed non-polar part of residue; $X_5$, molecular mass weighted exposed area; $X_6$, average deviation from the standard exposed area of residue; $X_7$, euclidian distance; $X_8$, secondary structure penalty; $X_9$, spacial distribution constraints (N,K Value). We randomly partition the dataset into a training set and a test set according to $7:3$, where the training set has 32,010 samples and the test set has 13,720 samples. In this instance, we build a linear regression model using the training data and then use the test data to assess the accuracy of the SSMEL's prediction under $K=[10,50,100]$. We recorded the mean squared prediction error (MSPE) and computation time (CT) of the CEL, DEL, and SSMEL under different numbers of the split in Table~\ref{real_analy1}. We can observe from Table~\ref{real_analy1} that when $K$ is close to 100, the computation time is greatly shortened and the MSPE of the SSMEL is very close to the CEL. Although the DEL succeeds in removing computational obstacles at large $K$, its MPSE is higher than that of the SSMEL and CEL.

\begin{table}[]
\centering
\tabcolsep=0.5cm
\renewcommand\arraystretch{1}
\caption{The mean of $(\hat{R}_{predict}-R_{true})^2$ and computation time under the different methods}\label{real_analy1}
\begin{tabular}{cccc}
\toprule
Method  &  $K$       &         MSPE            & CT(s) \\
\midrule
CEL       &  -----    &32.8624(48.2351)   &101.0488\\
DEL       &  10      &33.7463(50.2541)    &17.5427\\
             &  50      &41.2760(66.1366)    &7.3695 \\
             &  100    &52.8182(91.1082)    &2.5973 \\
SSMEL   &  10      &39.8627(57.9259)    &2.4526 \\
             &  50      &33.8907(50.1201)    &2.5971 \\
             &  100    &32.5892(47.7523)    &2.6080 \\
\bottomrule
\end{tabular}
\end{table}

\subsection{The United Stated airline dataset}
In this subsection, we use the SSMEL to analyze the United States airline dataset, which is publicly available on the American Statistical Association (ASA) website\footnote{\url{http://stat-computing.org/dataexpo/2009}}. This airline dataset is very large, with nearly 120 million records. Each record contains information on every commercial flight detail in the United States from October 1987 to April 2008. The dataset is partitioned into 22 files based on year, each file containing 13 continuous variables and 16 categorical variables. However, due to the massive size of the dataset, a typical personal computer may not have sufficient memory to load the full dataset for statistical analysis. In this paper, we concentrate on the analysis of the 13 continuous variables, and only 5 have missing rates less 10\%: ActualElapsedTime (actual elapsed time), CRSElapsedTime (scheduled elapsed time), Distance, DepDelay (departure) and ArrDelay (arrival delay). Therefore, we study these 5 variables. For more detailed information on the variables, refer to the ASA official website.

Due to these variables being so heavy-tailed that the existence of finite moments becomes questionable. Similar to \citet{Wu2021Subsampling}, we perform a signed-log-transformation: $\log|x|\cdot {\rm sign}(x)$ on these variables. For each transformed variable, we examine the mean, standard deviation, skewness, and kurtosis, denoted as $\mu$, $\sigma$, $\xi$, and $\kappa$. These statistics satisfy the following moment conditions:
\begin{equation*}
\mathbb{E}\big[g(X;\theta_{0})\big]=\mathbb{E}
\begin{pmatrix}
 \mu - X \\
\sigma^{2}-(X-\mu)^{2} \\
\xi - (X-\mu)^3/\sigma^3 \\
\kappa-(X-\mu)^4/\sigma^4
\end{pmatrix}
=0.
\end{equation*}

Due to the unacceptable size of this dataset and the extraordinarily extensive computing time, it can be hard to apply the empirical likelihood. We simulate distributed computing using this data. To simulate the distributed data situation, 22 parallel pools are created using the \texttt{parallel} package, and the data is calculated in one parallel pool each year. The data for each year is randomly partitioned into 5 subsets, totaling 110 subsets and the SSMEL is calculated using Algorithm~\ref{algo1} in Section 4. Additionally, to compare, we compute the DEL for the $K=110$ and report the results in Table~\ref{asa_data}. From Table~\ref{asa_data}, we can see that the estimators of the two methods are consistent for most variables, while the SSMEL has better computation time than the DEL.

\begin{table}[!htbp]
\centering
\caption{Estimators of five continuous variables after a signed-log transformation}\label{asa_data}
\resizebox{\textwidth}{!}{
\begin{tabular}{ccccccccccccc}
\toprule
\multirow{2}{*}{Method} &  & \multirow{2}{*}{Parameter} &  & Actual &  & CRS &  & \multirow{2}{*}{Distance} &  & \multirow{2}{*}{DepDelay} &  & \multirow{2}{*}{ArrDelay} \\
                   &  &                       &  & ElapsedTime &  & ElapsedTime &  &                     &  &                     &  &                     \\
\midrule
\multirow{5}{*}{DEL} 
                  &  & $\mu$      &  &4.6346&  &4.6457&  &6.2564&  &0.7949&  &0.3259\\
                  &  & $\sigma$  &  &0.5276&  &0.5154&  &0.7757&  &1.8847&  &2.3783\\
                  &  & $\xi$        &  &0.1741&  &0.2239&  &-0.1589&  &0.4074&  &0.1547\\
                  &  & $\kappa$  &  &2.6680&  &2.6548&  &2.7597&  &2.0858&  &1.6773\\
                  &  &CT(s)          &  &12784.84&  &15665.45&  &9936.08&  &12939.32 &  &12477.89   \\
\midrule
                  \multirow{5}{*}{SSMEL} 
                  &  & $\mu$      &  & 4.6389   &  &  4.6546  &  &6.2578 &  &0.7349     &  &0.2805  \\
                  &  & $\sigma$  &  & 0.5312   &  &  0.5209  &  &0.7792 &  &1.9461     &  &2.4031\\
                  &  & $\xi$        &  &0.1586    &  & 0.1394   &  &-0.1679&  &0.3536&  &0.1765\\
                  &  & $\kappa$  &  & 2.6751   &  & 2.6683   &  &2.7680&  &2.0015&  &1.6441\\
                  &  &CT(s)         &  & 992.46     &  &1431.15      &  &1002.33  &  &685.56  &  &1364.29\\
\bottomrule
\end{tabular}
}
\end{table}

\section{Conclusion}
In this paper, we propose a novel and straightforward methodology for calculating the empirical likelihood with massive data, which we refer to as split sample mean empirical likelihood (SSMEL). The approach uses split and compression techniques to overcome the challenges of empirical likelihood with massive data.  We show that the SSMEL preserves the statistical properties of empirical likelihood, making it suitable for parameter estimation and statistical inference. The effectiveness of our approach has been verified through both extensive simulation and real data analysis. Our method does not require parallel computation, which means it can be used on a wide range of computing devices and real-world applications. Additionally, to make it easier to process distributed data, we have developed a corresponding distributed algorithm for the SSMEL.

To conclude this paper, we discuss several intriguing avenues for future research. Initially, we focused solely on fixed dimensionality, but given the prevalence of high-dimensional massive data in real-world applications, expanding the approach to encompass cases where both $n$ and $p$ are large is crucial. Additionally, exploring the extension of this idea to more general $M$ estimators would be of significant interest. Finally, while we have developed a distributed dataset algorithm for the SSMEL, it is essentially a basic extension of the algorithm presented in \citet{tang2014algo}. Unfortunately, this approach incurs a significant communication cost. Therefore, exploring the possibility of designing a more efficient distributed algorithm for this purpose would be worthwhile.

\section*{Acknowledgments}
Chen's research was supported by the National Natural Science Foundation of China (12031016). Yang's research was supported by the National Social Science Foundation of China (22BTJ017).

\appendix
\section{Proofs}

\begin{lemma}\label{lem1}
Under the Assumptions~{\rm\protect\ref{A1}-\ref{A3}}, for any $\xi$ with $1/\alpha<\xi<1/2$, we have
$$
\sup_{\theta\in\bar{\Theta},\lambda\in\Lambda_{n},1\le k\le K}\left|\lambda^{T}\bar{g}^{(k)}(\theta)\right|\stackrel{p}{\longrightarrow}0,
$$
with probability tending to $1$ for all $\lambda\in\Lambda_{n}=\left\{\lambda:\Vert\lambda\Vert\le n^{-\xi}\right\}$, and $\Lambda_{n}\subset\hat{\Lambda}_{K}(\theta)$ for all $\theta\in\bar{\Theta}$.
\end{lemma}

\begin{pf}
By the Assumption~\protect\ref{A3} and the Markov inequality, we have
\begin{equation*}
\sup\limits_{\theta\in\bar{\Theta}}\left\Vert g(x_{i},\theta)\right\Vert= O_{p}\left(n^{1/\alpha}\right),    
\end{equation*}
and on the other hand,
\begin{align*}
\max\limits_{1\le k\le K}\sup\limits_{\theta\in\bar{\Theta}}\left\Vert\bar{g}^{(k)}(\theta)\right\Vert&=\max\limits_{1\le k\le K}\sup\limits_{\theta\in\bar{\Theta}}\left\Vert\frac{1}{m}\sum\limits_{i=1}\limits^{m}g(x_{i}^{(k)},\theta)\right\Vert \\
&\le\max\limits_{1\le k\le K}\sup\limits_{\theta\in\bar{\Theta}}\frac{1}{m}\sum\limits_{i=1}\limits^{m}\left\Vert g(x_{i}^{(k)},\theta)\right\Vert \\
&\le\max\limits_{1\le k\le K}\sup\limits_{\theta\in\bar{\Theta}}\frac{1}{m}\sum\limits_{i=1}\limits^{m}\max\limits_{1\le i\le m}\left\Vert g(x_{i}^{(k)},\theta)\right\Vert \\
&=\max\limits_{1\le i\le n}\sup\limits_{\theta\in\bar{\Theta}}\left\Vert g(x_{i},\theta)\right\Vert.
\end{align*}
So it is obvious that
\begin{equation*}
\max\limits_{1\le k\le K}\sup\limits_{\theta\in\bar{\Theta}}\left\Vert\bar{g}^{(k)}(\theta)\right\Vert=O_{p}\left(n^{1/\alpha}\right),   
\end{equation*}
then by the Cauchy-Schwarz inequality,
\begin{equation*}
\sup\limits_{\theta\in\bar{\Theta},\lambda\in\Lambda_{n},1\le k\le K}\left|\lambda^{T}\bar{g}^{(k)}(\theta)\right|\le n^{-\xi}\max\limits_{1\le k\le K}\sup\limits_{\theta\in\bar{\Theta}}\left\Vert \bar{g}^{(k)}(\theta)\right\Vert=O_{p}\left(n^{-\xi+1/\alpha}\right)\stackrel{p}{\longrightarrow}0.   
\end{equation*}
\end{pf}

\begin{lemma}\label{lem2}
Under the Assumptions~{\rm\protect\ref{A1}-\ref{A4}}, if $\hat{\theta}\in\bar{\Theta}$, $\hat{\theta}\stackrel{p}{\longrightarrow}\theta_0$, and $K^{-1}\sum_{k=1}^{K}\bar{g}^{(k)}(\hat{\theta})=O_{p}\left(n^{-1/2}\right)$, then with probability tending to $1$, $\hat{\lambda}=\arg\max_{\lambda\in\hat{\Lambda}_{n}(\hat{\theta})}\ell_{S}(\lambda,\hat{\theta})$ exists, and $\max_{\lambda\in\hat{\Lambda}_{n}(\hat{\theta})}\ell_{S}(\lambda,\hat{\theta})\le O_{p}\left(n^{-1}\right)$.
\end{lemma}

\begin{pf}
The existence of $\hat{\lambda}\in\Lambda_n$ follows the statement in \citet{newey2004higher} by noting from Lemma~\protect\ref{lem1}, $\max_{1\le k\le K}\left|\lambda^{T}\bar{g}^{(k)}(\theta)\right|\stackrel{p}{\longrightarrow}0$ for $\lambda\in\Lambda_{n}$.  Then by the Taylor expansion around $\lambda=0$, 
\begin{equation*}
K\cdot \ell_{S}(\hat{\theta},\hat{\lambda})=\sum\limits_{k=1}\limits^{K}\hat{\lambda}^{T}\bar{g}^{(k)}(\hat{\theta})-\frac{1}{2}\hat{\lambda}^{T}\left[\sum\limits_{k=1}\limits^{K}\big\{1+\dot{\lambda}^{T}\bar{g}^{(k)}(\hat{\theta})\big\}^{-2}\bar{g}^{(k)}(\hat{\theta})\bar{g}^{(k)}(\hat{\theta})^{T}\right]\hat{\lambda},
\end{equation*}
where $\dot{\lambda}$ satisfy $\Vert \dot{\lambda}\Vert \le \Vert\hat{\lambda}\Vert$.
By Lemma~\protect\ref{lem1}, $\left(1+\dot{\lambda}\bar{g}^{(k)}(\hat{\theta})\right)^{-2}>1/2$ for all $k$ with probability tending to 1. In addition, by the weak law of large numbers, as $n\rightarrow\infty$, 
\begin{equation*}
\left\Vert\frac{1}{K}\sum_{k=1}^{K}\bar{g}^{(k)}(\hat{\theta})\bar{g}^{(k)}(\hat{\theta})^{T}-\frac{1}{m}\Omega\right\Vert\stackrel{p}{\longrightarrow}0. 
\end{equation*}
Because $\hat{\lambda}$ is the maximizer, with probability tending to 1,
\begin{equation}\label{l_bound}
0=\ell_{S}(\hat{\theta},0)\le \ell_{S}(\hat{\theta},\hat{\lambda})\le\Vert\hat{\lambda}\Vert\left\Vert\frac{1}{K}\sum\limits_{k=1}\limits^{K}\bar{g}^{(k)}(\hat{\theta})\right\Vert-\frac{c}{4}\Vert\hat{\lambda}\Vert^2.
\end{equation}
This concludes $\Vert\hat{\lambda}\Vert=O_{p}\left(n^{-1/2}\right)$ because $\left\Vert K^{-1}\sum_{k=1}^{K}\bar{g}^{(k)}(\hat{\theta}) \right\Vert = O_{p}\left(n^{-1/2}\right)$. Since $\xi\le1/2$, we have $\hat{\lambda}\in\hat{\Lambda}_{K}(\hat{\theta})$ with probability tending to 1. And this is easy to get $\max_{\lambda\in\hat{\Lambda}_{K}(\hat{\theta})}\ell_{S}(\lambda,\hat{\theta})\le O_{p}\left(n^{-1}\right)$ from Equation~\protect(\ref{l_bound}).
\end{pf}

\begin{lemma}\label{lem3}
Under the Assumptions~{\rm\protect\ref{A1}-\ref{A4}}, we have $\left\Vert\frac{1}{K}\sum_{k=1}^{K}\bar{g}^{(k)}(\hat{\theta})\right\Vert=O_{p}\left(n^{-1/2}\right)$.
\end{lemma}

\begin{pf}
It is worth noting that
\begin{equation*}
\frac{1}{K}\sum\limits_{k=1}\limits^{K}\bar{g}^{(k)}(\hat{\theta})=\frac{1}{K}\sum\limits_{k=1}\limits^{K}\frac{1}{m}\sum\limits_{i=1}\limits^{m}g(x_{i}^{(k)},\hat{\theta})=\frac{1}{n}\sum\limits_{i=1}\limits^{n}g(x_{i},\hat{\theta}).
\end{equation*}

Let $\bar{g}(\hat{\theta})=K^{-1}\sum_{k=1}^{K}\bar{g}^{(k)}(\hat{\theta})=n^{-1}\sum_{i=1}^{n}g(x_{i},\hat{\theta})$, and for $\xi$ in Lemma~\ref{lem1}, $\tilde{\lambda}=n^{-\xi}\bar{g}(\hat{\theta})/\Vert\bar{g}(\hat{\theta})\Vert$. By Lemma~\ref{lem1}, $\max_{1\le k\le K}|\tilde{\lambda}^{T}\bar{g}^{(k)}(\hat{\theta})|\stackrel{p}{\longrightarrow}0$ and $\tilde{\lambda}\in\hat{\Lambda}_{K}(\hat{\theta})$ with probability tending to 1. Also, by the Cauchy-Schwarz inequality and the weak law of large numbers, 
\begin{equation*}
\frac{1}{K}\sum_{k=1}^{K}\bar{g}^{(k)}(\hat{\theta})\bar{g}^{(k)}(\hat{\theta})^{T}\le\left(\frac{1}{K}\sum_{k=1}^{K}\Vert\bar{g}^{(k)}(\hat{\theta})\Vert^2\right)I\stackrel{p}{\longrightarrow}CI,
\end{equation*}
so the largest eigenvalue of $\sum_{k=1}^{K}\bar{g}^{(k)}(\hat{\theta})\bar{g}^{(k)}(\hat{\theta})^{T}/K$ is bounded above with probability tending to 1. By Taylor expansion, it holds with probability tending to 1,
\begin{align*}
\ell_{S}(\hat{\theta},\tilde{\lambda})
&=\frac{1}{K}\sum\limits_{k=1}\limits^{K}\tilde{\lambda}^{T}\bar{g}^{(k)}(\hat{\theta})-\frac{1}{2}\tilde{\lambda}^{T}\left[\frac{1}{K}\sum\limits_{i=1}\limits^{K}\left\{1+\dot{\lambda}^{T}\bar{g}^{(k)}(\hat{\theta})\right\}^{-2}\bar{g}^{(k)}(\hat{\theta})\bar{g}^{(k)}(\hat{\theta})^{T}\right]\tilde{\lambda}\\
&\ge n^{-\xi}\left\Vert\bar{g}(\hat{\theta})\right\Vert-Cn^{-2\xi}\left\{1+o_{p}(1)\right\}.
\end{align*}
where $\Vert \dot{\lambda}\Vert \le\Vert\tilde{\lambda}\Vert$. By the Lindeberg-L\'{e}vy central limit theorem, the hypotheses of Lemma~\ref{lem2} are satisfied by $\hat{\theta}=\theta_0$. By $\hat{\theta}$ and $\hat{\lambda}$ being a saddle point, this equation and Lemma~\ref{lem2} give
\begin{equation}\label{g_bound}
n^{-\xi}\left\Vert\bar{g}(\hat{\theta})\right\Vert-Cn^{-2\xi}\le \ell_{S}(\hat{\theta},\tilde{\lambda})\le \max\limits_{\lambda\in\hat{\Lambda}_{K}(\hat{\theta})}\ell_{S}(\hat{\theta},\lambda) \le O_{p}\left(n^{-1}\right).
\end{equation}
This gives
\begin{equation*}
\left\Vert \bar{g}(\hat{\theta})\right\Vert\le O_{p}\left(n^{\xi-1}\right)+Cn^{-\xi} = O_{p}\left(n^{-\xi}\right).
\end{equation*}
For any $\varepsilon_{n}\rightarrow0$, let $\lambda^{*}=\varepsilon_{n}\bar{g}(\hat{\theta})$, then $\lambda^{*}=o_{p}\left(n^{-\xi}\right)$ and
$\lambda^{*}\in \Lambda_{n}$ with probability tending to 1. Thus we can obtain
\begin{equation*}
\epsilon_n\left\Vert\bar{g}(\hat{\theta})\right\Vert^2-C\epsilon_{n}^{2}\le O_{p}\left(n^{-1}\right).
\end{equation*}
Then $\epsilon_n\Vert\bar{g}(\hat{\theta})\Vert^2=O_{p}\left(n^{-1}\right)$. Notice that we can select arbitrary slow $\varepsilon_{n}\rightarrow0$, following a standard result form probability theory, that if $\varepsilon_{n}Y_{n}=O_{p}\left(n^{-1}\right)$, for all $\varepsilon_n\rightarrow0$, then $Y_{n}=O_{p}\left(n^{-1}\right)$. So, we have $\Vert\bar{g}(\hat{\theta})\Vert=O_{p}\left(n^{-1/2}\right)$.
\end{pf}

\begin{lemma}\label{lem4}
Under the Assumptions~{\rm\protect\ref{A1}-\ref{A5}}, we have 
\begin{equation*}
\hat{\theta}_{S}\stackrel{p}{\longrightarrow}\theta_0,
\end{equation*}
\begin{equation*}
K^{-1}\sum_{k=1}^{K}\bar{g}^{(k)}(\hat{\theta}_{S})=O_{p}\left(n^{-1/2}\right),
\end{equation*}
and $\hat{\lambda}_{S}=\arg\max_{\lambda\in\hat{\Lambda}_{K}(\hat{\theta}_{S})}\ell_{S}(\hat{\theta}_{S},\lambda)$
exists with probability tending to $1$, and $\hat{\lambda}_{S}=O_{p}\left(n^{-1/2}\right)$.
\end{lemma}

\begin{pf}
Let $g(\theta)=\mathbb{E}[g(X,\theta)]$, by Lemma~\ref{lem3}, $\bar{g}(\hat{\theta}_{S})\stackrel{p}{\longrightarrow}{0}$, and by the uniform weak law of large numbers, $\sup_{\theta\in\bar{\Theta}}\Vert\bar{g}(\theta)-g(\theta)\Vert\stackrel{p}{\longrightarrow}{0}$ and $g(\theta)$ is continuous. By the triangle inequality,
\begin{equation*}
\Vert \bar{g}(\hat{\theta}_{S})\Vert-\Vert g(\hat{\theta}_{S})\Vert\le\Vert\bar{g}(\hat{\theta}_{S})-g(\hat{\theta}_{S})\Vert\le\sup_{\theta\in\bar{\Theta}}\left\Vert\bar{g}(\theta)-g(\theta)\right\Vert.
\end{equation*}
We have $g(\hat{\theta}_{S})\stackrel{p}{\longrightarrow}{0}$. Since $g(\theta)=0$ has a unique zero at $\theta_0$, $\Vert g(\theta)\Vert$ must be bounded away from zero outside any neighborhood of $\theta_0$. Therefore, $\hat{\theta}_{S}$ must be inside any neighborhood of $\theta_0$ with probability tending to $1$, i.e. $\hat{\theta}_{S}\stackrel{p}{\longrightarrow}{\theta_0}$, giving the first conclusion. The second conclusion follows by Lemma~\protect\ref{lem3}. And by the first two conclusions, the hypotheses of Lemma~\ref{lem2} are satisfied, so the last conclusion follows from Lemma~\protect\ref{lem2}.
\end{pf}

\begin{lemma}\label{lem5}
Under the Assumptions~{\rm\protect\ref{A1}-\ref{A5}}, $\hat{\theta}_{S}$ and $\hat{\lambda}_{S}$ satisfy
\begin{align*}
Q_{1K}\left(\hat{\theta}_{S},\hat{\lambda}_{S}\right)=0,\quad Q_{2K}\left(\hat{\theta}_{S},\hat{\lambda}_{S}\right)=0,
\end{align*}
where
\begin{align*}
		Q_{1K}(\theta,\lambda)&=\frac{1}{K}\sum\limits_{k=1}^{K}\frac{1}{1+\lambda^{T}\bar{g}^{(k)}(\theta)}\bar{g}^{(k)}(\theta),\\
		Q_{2K}(\theta,\lambda)&=\frac{1}{K}\sum\limits_{k=1}^{K}\frac{1}{1+\lambda^{T}\bar{g}^{(k)}(\theta)}\Big(\frac{\partial \bar{g}^{(k)}(\theta)}{\partial\theta}\Big)^{T}\lambda.
\end{align*}
\end{lemma}

\begin{pf}
The conclusion can be obtained from Lemma~\ref{lem1} and Lemma~\ref{lem4}, more details refer the proof of Lemma 1 of \citet{qin1994} or the proof of Theorem 3.2 of \citet{newey2004higher}.
\end{pf}

\begin{pot1}
By Lemma~\ref{lem5}, we have
\begin{align*}
		Q_{1K}(\theta_S,\lambda_S)&=\frac{1}{K}\sum\limits_{k=1}^{K}\frac{1}{1+\lambda_{S}^{T}\bar{g}^{(k)}(\theta_S)}\bar{g}^{(k)}(\theta_{S})=0,\\
		Q_{2K}(\theta_S,\lambda_S)&=\frac{1}{K}\sum\limits_{k=1}^{K}\frac{1}{1+\lambda^{T}_{S}\bar{g}^{(k)}(\theta_S)}\left(\frac{\partial \bar{g}^{(k)}(\theta_S)}{\partial\theta}\right)^{T}\lambda_{S}=0.
\end{align*}
As $n\rightarrow\infty$,

\begin{align*}
\frac{\partial Q_{1K}(\theta_{0},0)}{\partial\theta}&=\frac{1}{K}\sum\limits_{k=1}^{K}\frac{\partial \bar{g}^{(k)}(\theta_{0})}{\partial\theta^{T}}
=\frac{1}{K}\sum\limits_{k=1}^{K}\frac{1}{m}\sum\limits_{i=1}^{m}\frac{\partial g(x_{i}^{(k)},\theta_{0})}{\partial\theta^{T}}\\
&=\frac{1}{n}\sum\limits_{i=1}\limits^{n}\frac{\partial g(x_{i},\theta_0)}{\partial\theta^{T}}\stackrel{p}{\longrightarrow}\mathbb{E}\left(\frac{\partial g}{\partial\theta^{T}}\right)=G,\\
\end{align*}
\begin{align*}
\frac{\partial Q_{1K}(\theta_{0},0)}{\partial\lambda^{T}}&=-\frac{1}{K}\sum\limits_{k=1}^{K}\bar{g}^{(k)}(\theta_{0})\bar{g}^{(k)}(\theta_{0})^{T}\\
&=-\frac{1}{K}\sum\limits_{k=1}^{K}\frac{1}{m^{2}}\left(\sum\limits_{i=1}^{m}g(x_{i}^{(k)},\theta_{0})\sum\limits_{i=1}^{m}g(x_{i}^{(k)},\theta_{0})^{T}\right)\\
&=-\frac{1}{m}\frac{1}{n}\sum\limits_{i=1}\limits^{n}g(x_{i},\theta_0)g(x_{i},\theta_0)^{T}\\
&-\frac{1}{n}\sum\limits_{k=1}\limits^{K}\left(\frac{2}{m}\sum\limits_{i\neq s}^{m}g(x_{i}^{(k)},\theta_0)g(x_{s}^{(k)},\theta_0)^{T}\right)\\
&\stackrel{p}{\longrightarrow}-\frac{1}{m}\mathbb{E}\left(gg^{T}\right)+o_{p}(1)=-\frac{1}{m}\Omega+o_{p}(1),\\
\end{align*}
\begin{align*}
\frac{\partial Q_{2K}(\theta_{0},0)}{\partial\lambda^{T}}&=\frac{1}{K}\sum\limits_{k=1}^{K}\left(\frac{\partial \bar{g}^{(k)}(\theta_{0})}{\partial\theta^{T}}\right)^{T}=\frac{1}{K}\sum\limits_{k=1}^{K}\frac{1}{m}\sum\limits_{i=1}^{m}\frac{\partial g(x_{i}^{(k)},\theta_{0})}{\partial\theta^{T}}^{T}\\
&=\frac{1}{n}\sum\limits_{i=1}\limits^{n}\frac{\partial g(x_{i},\theta_0)^{T}}{\partial\theta^{T}}\stackrel{p}{\longrightarrow}\mathbb{E}\left(\frac{\partial g}{\partial\theta^{T}}\right)^{T}=G^{T}.\\
\end{align*}
By the Taylor expansion around $(\theta_0,0)$, we can show

\begin{align*}
0&=Q_{1K}\left(\hat{\theta}_{S},\hat{\lambda}_{S}\right)=Q_{1K}(\theta_{0},0)+\frac{\partial Q_{1K}(\theta_{0},0)}{\partial\theta}\left(\hat{\theta}_{S}-\theta_{0}\right)+\frac{\partial Q_{1K}(\theta_{0},0)}{\partial\lambda^{T}}\hat{\lambda}_{S}+o_{p}\left(\delta_{K}\right),\\
0&=Q_{2K}\left(\hat{\theta}_{S},\hat{\lambda}_{S}\right)=Q_{2K}(\theta_{0},0)+\frac{\partial Q_{2K}(\theta_{0},0)}{\partial\theta}\left(\hat{\theta}_{S}-\theta_{0}\right)+\frac{\partial Q_{2K}(\theta_{0},0)}{\partial\lambda^{T}}\hat{\lambda}_{S}+o_{p}\left(\delta_{K}\right),
\end{align*}
where both $Q_{2K}(\theta_{0},0)$ and $\partial Q_{2K}(\theta_{0},0)/\partial\theta$ are 0, $\delta_{K}=\left\Vert\hat{\theta}_{S}-\theta_{0}\right\Vert+\left\Vert\hat{\lambda}_{S}\right\Vert$, so

\begin{align*}
0&=\frac{\partial Q_{2K}(\theta_{0},0)}{\partial\lambda^{T}}\times\left(-\frac{\partial Q_{1K}(\theta_{0},0)}{\partial\lambda^{T}}\right)^{-1}\times\left[Q_{1K}(\theta_{0},0)+\frac{\partial Q_{1K}(\theta_{0},0)}{\partial\theta}\left(\hat{\theta}_{S}-\theta_{0}\right)+o_{p}\left(\delta_{K}\right)\right]\\
&=G^{T}\left(\frac{1}{m}\Omega\right)^{-1}Q_{1K}(\theta_{0},0)+G^{T}\left(\frac{1}{m}\Omega\right)^{-1}G\left(\hat{\theta}_{S}-\theta_{0}\right)+o_{p}\left(mn^{-1/2}\right).
\end{align*}
It means
\begin{align}\label{thetadiff}
\hat{\theta}_{S}-\theta_{0}&=-\Sigma G^{T}\Omega^{-1} Q_{1K}(\theta_{0},0)+o_{p}\left(n^{-1/2}\right).
\end{align}
Because $-\sqrt{n}\Omega^{-1/2}Q_{1K}(\theta_0,0)$ converges to standard multivariate normal distribution, i.e. $-\sqrt{n}\Omega^{-1/2}Q_{1K}(\theta_0,0)=-\sqrt{n}\Omega^{-1/2}\bar{g}(\theta_0)\stackrel{d}{\longrightarrow}N(0,I)$, therefore as $n\rightarrow\infty$, we have
\begin{equation*}
\sqrt{n}\left(\hat{\theta}_{S}-\theta_0\right)\stackrel{d}{\longrightarrow}N(0,\Sigma).
\end{equation*}
\end{pot1}

\begin{poc1}
By Equation~\protect(\ref{thetadiff}), we denote $\Sigma G^{T}\Omega=B$, therefore,
\begin{equation}\label{thetadiff2}
\hat{\theta}_{S}-\theta_0 = -BQ_{1K}(\theta_0,0)+o_{p}\left(n^{-1/2}\right).
\end{equation}
To derive the upper bound for MSE of $\hat{\theta}_{S}$, we first take the following algebraic calculation:
\begin{equation*}
\mathbb{E}\left[\left\Vert-BQ_{1K}(\theta_0,0)\right\Vert^2\right]=\mathbb{E}\left[\left\Vert-B\frac{1}{K}\sum\limits_{k=1}^{K}\bar{g}^{(k)}(\theta_0)\right\Vert^2\right]=\mathbb{E}\left[\left\Vert-B\frac{1}{n}\sum\limits_{i=1}^{n}g(x_i,\theta_0)\right\Vert^2\right].
\end{equation*}
By the proof of Lemma A.1. from \citet{del2022}, we have
\begin{equation}\label{mse_bound1}
\mathbb{E}\left[\left\Vert-B\frac{1}{n}\sum\limits_{i=1}^{n}g(x_i,\theta_0)\right\Vert^2\right] = \frac{tr(\Sigma)}{n}.
\end{equation}
Hence, by Equation~\protect(\ref{thetadiff2}) and (\ref{mse_bound1}), the MSE of SSMEL estimator $\hat{\theta}_{S}$ is calculated as follows:
\begin{small}
\begin{align*}
\mathbb{E}\left[\left\Vert\hat{\theta}_{S}-\theta_0\right\Vert^2\right]&=\mathbb{E}\left[\left\Vert-B\frac{1}{K}\sum\limits_{k=1}^{K}\bar{g}^{(k)}(\theta_0)+o_{p}\left(n^{-1/2}\right)\right\Vert^2\right]\\
&=\mathbb{E}\left[\left\Vert-B\frac{1}{n}\sum\limits_{i=1}^{n}g(x_i,\theta_0)+o_{p}\left(n^{-1/2}\right)\right\Vert^2\right]\\
&=\mathbb{E}\left[\left\Vert B\frac{1}{n}\sum\limits_{i=1}^{n}g(x_i,\theta_0)+o_{p}\left(n^{-1/2}\right)\right\Vert^2\right]\\
&=\mathbb{E}\left[\left(B\frac{1}{n}\sum\limits_{i=1}^{n}g(x_i,\theta_0)+o_{p}\left(n^{-1/2}\right)\textbf{c}\right)^{T}\left(B\frac{1}{n}\sum\limits_{i=1}^{n}g(x_i,\theta_0)+o_{p}\left(n^{-1/2}\right)\textbf{c}\right)\right]\\
&=\mathbb{E}\left[\left\Vert B\frac{1}{n}\sum\limits_{i=1}^{n}g(x_i,\theta_0)\right\Vert^2 + o_{p}\left(n^{-1}\right)\textbf{c}^{T}\textbf{c}+2o_{p}\left(n^{-1/2}\right)\textbf{c}^{T}B\frac{1}{n}\sum\limits_{i=1}^{n}g(x_i,\theta_0)\right]\\
&\le \mathbb{E}\left[\left\Vert B\frac{1}{n}\sum\limits_{i=1}^{n}g(x_i,\theta_0)\right\Vert^2 + o_{p}\left(n^{-1}\right)\textbf{c}^{T}\textbf{c}+2o_{p}\left(n^{-1/2}\right)\sqrt{\textbf{c}^{T}\textbf{c}}\sqrt{\left\Vert B\frac{1}{n}\sum\limits_{i=1}^{n}g(x_i,\theta_0)\right\Vert^2}\right]\\
&\le \mathbb{E}\left[\left\Vert B\frac{1}{n}\sum\limits_{i=1}^{n}g(x_i,\theta_0)\right\Vert^2\right]+o\left(n^{-1}\right) + o\left(n^{-1/2}\right)\sqrt{\mathbb{E}\left[\left\Vert B\frac{1}{n}\sum\limits_{i=1}^{n}g(x_i,\theta_0)\right\Vert^2\right]}\\
&=\frac{tr(\Sigma)}{n} + o\left(n^{-1}\right)+o\left(n^{-1/2}\right)\sqrt{\frac{tr(\Sigma)}{n}}\\
&=\frac{tr(\Sigma)}{n} + o\left(n^{-1}\right) + o\left(n^{-1/2}\right)O\left(n^{-1/2}\right)\\
&= \frac{tr(\Sigma)}{n} + o\left(n^{-1}\right),
\end{align*}
\end{small}
where \textbf{c} is an arbitrary $p$-dimensional constant vector.
\end{poc1}

\begin{pot2}
The split sample mean empirical likelihood ratio test statistic is
\begin{equation*}
\mathcal{W}(\theta_0)=2\left\{\sum\limits_{k=1}^{K}\log\left[1+\lambda_{0}^{T}\bar{g}^{(k)}(\theta_0)\right]-\sum\limits_{k=1}^{K}\log\left[1+\hat{\lambda}_{S}^{T}\bar{g}^{(k)}(\hat{\theta}_{S})\right]\right\}.
\end{equation*}
By Lemma~\protect\ref{lem5}, we have $Q_{1K}(\hat{\theta}_{S},\hat{\lambda}_{S})=0$, and by Tylor expansion
\begin{align}\label{labda}
\hat{\lambda}_{S}&=\left[\frac{1}{K}\sum\limits_{k=1}\limits^{K}\bar{g}^{(k)}(\hat{\theta}_{S})\bar{g}^{(k)}(\hat{\theta}_{S})^{T}\right]^{-1}\left(\sum\limits_{k=1}\limits^{K}\bar{g}^{(k)}(\hat{\theta}_{S})\right)+o_{p}\left(1\right) \nonumber\\
&=\left(\frac{1}{m}\Omega\right)^{-1}Q_{1K}(\hat{\theta}_{S},0)+o_{p}\left(1\right).
\end{align}
Also by the Taylor expansion and  Euqation~\protect(\ref{thetadiff}),
\begin{align}\label{Q1K}
Q_{1K}(\hat{\theta}_{S},0)&=Q_{1K}(\theta_0,0)+\frac{\partial Q_{1K}(\theta_0,0)}{\partial\theta^{T}}\left(\hat{\theta}_{S}-\theta_0\right)+o_{p}\left(1\right)\nonumber,\\
&=Q_{1K}(\theta_0,0)+G\left(\hat{\theta}_{S}-\theta_0\right)+o_{p}\left(1\right)\nonumber\\
&=Q_{1K}(\theta_0,0)-G\Sigma G^{T}\Omega^{-1}Q_{1K}(\theta_{0},0)+o_{p}\left(1\right).
\end{align}

Further Taylor expansion for $\ell_{S}(\hat{\theta}_{S},\hat{\lambda}_{S})$, and by Equation~\protect(\ref{labda}) and \protect(\ref{Q1K}) we have
\begin{align*}
2\ell_{S}(\hat{\theta}_{S},\hat{\lambda}_{S})&=2\sum\limits_{k=1}^{K}\log\left[1+\hat{\lambda}_{S}^{T}\bar{g}^{(k)}(\hat{\theta}_{S})\right]\\
&=2\sum\limits_{k=1}\limits^{K}\hat{\lambda}^{T}_{S}\bar{g}^{(k)}(\hat{\theta}_{S})-\sum\limits_{k=1}\limits^{K}\left[\hat{\lambda}^{T}_{S}\bar{g}^{(k)}(\hat{\theta}_{S})\right]^{2}+o_{p}\left(1\right)\\
&=KQ_{1K}^{T}(\hat{\theta}_{S},0)\left[\frac{1}{K}\sum\limits_{k=1}\limits^{K}\bar{g}^{(k)}(\hat{\theta}_{S})\bar{g}^{(k)}(\hat{\theta}_{S})^{T}\right]^{-1}Q_{1K}(\hat{\theta}_{S},0)+o_{p}\left(1\right)\\
&=KQ_{1K}^{T}(\hat{\theta}_{S},0)\left(\frac{1}{m}\Omega\right)^{-1}Q_{1K}(\hat{\theta}_{S},0)+o_{p}\left(1\right)\\
&=nQ^{T}_{1K}(\theta_0,0)\Omega^{-1}\left(I-G\Sigma G^{T}\Omega^{-1}\right)Q_{1K}(\theta_0,0)+o_{p}\left(1\right).
\end{align*}
Under $H_{0}$ is ture, similarly
$$
\lambda_0=\Omega^{-1}Q_{1K}(\theta_0,0)+o_{p}(1),\quad and\quad 2\ell_{S}(\theta_0,\lambda_0)=nQ^{T}_{1K}(\theta_0,0)\Omega^{-1}Q_{1K}(\theta_0,0)+o_{p}(1).
$$
Thus
\begin{align*}
\mathcal{W}(\theta_0)&=nQ^{T}_{1K}(\theta_0,0)\left[\Omega^{-1}-\Omega^{-1}(I-G\Sigma G^{T}\Omega^{-1})\right]Q^{T}_{1K}(\theta_0,0)+o_{p}(1)\\
&=nQ^{T}_{1K}(\theta_0,0)\Omega^{-1}G\Sigma G^{T}\Omega^{-1}Q_{1K}(\theta_0,0)+o_{p}(1)\\
&=\left[\Omega^{-1/2}\sqrt{n}Q_{1K}(\theta_0,0)\right]^{T}\left[\Omega^{-1/2}G\Sigma G^{T}\Omega^{-1/2}\right]\left[\Omega^{-1/2}\sqrt{n}Q_{1K}(\theta_0,0)\right]+o_{p}(1).
\end{align*}
Note that $\Omega^{-1/2}\sqrt{n}Q_{1K}(\theta_0,0)$ converges to a standard multivariate normal distribution and that $\Omega^{-1/2}G\Sigma G^{T}\Omega^{-1/2}$ is symmetric idempotent, with trace equal to $p$. Hence the SSMEL ratio test statistic $\mathcal{W}(\theta_0)$ converges to $\chi_{p}^{2}$.
\end{pot2}

\begin{poc2}
Through the Taylor expansion, we have
\begin{align*}
\mathcal{W}(\phi_0)&=2\ell_{S}\left(\phi_0,\hat{\gamma}(\phi_0)\right)-2\ell_{S}\left(\hat{\phi},\hat{\gamma}\right)\\
&=\left[\Omega^{1/2}\sqrt{n}Q_{1K}(\phi_0,0)\right]^{T}\Omega^{-1/2}\\
&\times\left\{ G\Sigma^{-1}G^{T}-\left(\mathbb{E}\frac{\partial g}{\partial\phi}\right)\left[\left(\mathbb{E}\frac{\partial g}{\partial\phi}\right)^{T}\Omega^{-1}\left(\mathbb{E}\frac{\partial g}{\partial\phi}\right)\right]^{-1}\left(\mathbb{E}\frac{\partial g}{\partial\phi}\right)^{T}\right\}\\
&\times\Omega^{-1/2}\left[\Omega^{1/2}\sqrt{n}Q_{1K}(\phi_0,0)\right]+o_{p}(1).\\
\end{align*}
As a result of \citet{rao}, we only need to show that
\begin{align*}
\Delta &:=G\Sigma^{-1}G^{T}\\
&\ge \left(\mathbb{E}\frac{\partial g}{\partial\phi}\right)\left[\left(\mathbb{E}\frac{\partial g}{\partial\phi}\right)^{T}\Omega^{-1}\left(\mathbb{E}\frac{\partial g}{\partial\phi}\right)\right]^{-1}\left(\mathbb{E}\frac{\partial g}{\partial\phi}\right)^{T}.\\
\end{align*}
In fact,
\begin{align*}
\Delta &:=G\Sigma^{-1}G^{T}\\
&\ge \left(\mathbb{E}\frac{\partial g}{\partial\phi},\mathbb{E}\frac{\partial g}{\partial\gamma}\right)
\begin{pmatrix}
\left[\left(\mathbb{E}\frac{\partial g}{\partial\phi}\right)^{T}\Omega^{-1}\left(\mathbb{E}\frac{\partial g}{\partial\phi}\right)\right]^{-1}\left(\mathbb{E}\frac{\partial g}{\partial\phi}\right)^{T} & 0 \\
0 & 0 \\
\end{pmatrix}
\begin{pmatrix}
\mathbb{E}\left(\frac{\partial g}{\partial\phi}\right)^{T}\\
\mathbb{E}\left(\frac{\partial g}{\partial\gamma}\right)^{T}
\end{pmatrix}\\
&=\left(\mathbb{E}\frac{\partial g}{\partial\phi}\right)\left[\left(\mathbb{E}\frac{\partial g}{\partial\phi}\right)^{T}\Omega^{-1}\left(\mathbb{E}\frac{\partial g}{\partial\phi}\right)\right]^{-1}\left(\mathbb{E}\frac{\partial g}{\partial\phi}\right)^{T}.
\end{align*}
Thus $\mathcal{W}(\phi_0)\rightarrow\chi^{2}_{[r-(p-q)-(r-p)]}=\chi^2_{q}$.
\end{poc2}
\bibliographystyle{elsarticle-harv} 
\bibliography{ref}





\end{document}